\documentclass[showpacs,nofootinbib,preprintnumbers,prd,superscriptaddress,twocolumn]{revtex4}

\usepackage{amsmath}
\usepackage{amsfonts}
\usepackage{amssymb}
\usepackage{graphicx}
\usepackage[titletoc]{appendix}
\usepackage{color}
\usepackage{hyperref}
\usepackage{cleveref}
\usepackage[rightcaption]{sidecap}
\usepackage{subfigure}
\usepackage{comment}
\usepackage{tensor}
\usepackage{longtable}
\usepackage{graphicx}

\usepackage{mathtools}

\usepackage{dcolumn}

\usepackage{array}
\usepackage{ctable}
\usepackage{multirow}
\usepackage{siunitx}
\usepackage{longtable}
\usepackage{tabularx}
\usepackage{booktabs}

\newcommand{\gt}{>}
\newcommand{\lt}{<}

\usepackage{longtable}
\graphicspath{{Graphics/}}

\def\be{\begin{equation}}
\def\ee{\end{equation}}
\def\bea{\begin{eqnarray}}
\def\eea{\end{eqnarray}}

\definecolor{vividviolet}{rgb}{0.62, 0.0, 1.0}
\definecolor{amaranth}{rgb}{0.9, 0.17, 0.31}
\definecolor{palatinateblue}{rgb}{0.15, 0.23, 0.89}
\definecolor{brightpink}{rgb}{1.0, 0.0, 0.5}
\definecolor{cornflowerblue}{rgb}{0.39, 0.58, 0.93}
\definecolor{deepcarminepink}{rgb}{0.94, 0.19, 0.22}
\definecolor{radicalred}{rgb}{1.0, 0.21, 0.37}

\hypersetup{ linktoc=all,
    colorlinks, linkcolor={palatinateblue},
    citecolor={brightpink}, urlcolor={amaranth}
}

\begin{document}

\title{Phase-space analysis of dark energy models in non-minimally coupled theories of gravity}

\author{Youri Carloni}
\email{youri.carloni@unicam.it}
\affiliation{Universit\`a di Camerino, Via Madonna delle Carceri, Camerino, 62032, Italy.}
\affiliation{INAF - Osservatorio Astronomico di Brera, Milano, Italy.}
\affiliation{Istituto Nazionale di Fisica Nucleare (INFN), Sezione di Perugia, Perugia, 06123, Italy.}

\author{Orlando Luongo}
\email{orlando.luongo@unicam.it}
\affiliation{Universit\`a di Camerino, Via Madonna delle Carceri, Camerino, 62032, Italy.}
\affiliation{INAF - Osservatorio Astronomico di Brera, Milano, Italy.}
\affiliation{SUNY Polytechnic Institute, 13502 Utica, New York, USA.}
\affiliation{Istituto Nazionale di Fisica Nucleare (INFN), Sezione di Perugia, Perugia, 06123, Italy.}
\affiliation{Al-Farabi Kazakh National University, Al-Farabi av. 71, 050040 Almaty, Kazakhstan.}

\begin{abstract}
We analyze scalar field dark energy models minimally and non-minimally coupled to gravity, postulating that a Yukawa-like interacting term is \emph{in form} equivalent for general relativity, teleparallel and symmetric-teleparallel theories. Our analysis is pursued within two scalar field representations, where a quintessence and phantom pictures are associated with quasiquintessence and quasiphantom exotic fields. In the latter, we suggest how the phion-pressure can be built up without exhibiting a direct kinetic term.
Accordingly, the stability analysis reveals that this quasiquintessence field provides a viable description of the universe  indicating, when minimally coupled, how to unify dark energy and dark matter by showing an attractor point where $w_{\phi}=0$. Conversely,  in the non-minimally coupling, the alternative field only leaves an attractor where dark energy dominates, mimicking \emph{de facto} a cosmological constant behavior. A direct study is conducted comparing the standard case with the alternative one, overall concluding that the behavior of quintessence is well established across all the gravity scenarios. However, considering the phantom field non-minimal coupled to gravity, the results are inconclusive for power-law potentials in Einstein theory, and for the inverse square power (ISP) potential in both teleparallel and symmetric-teleparallel theories. Finally, we study the growth of matter perturbations and establish that only the fifth power and quadratic potentials, when used to describe quasiphantom field minimally coupled to gravity, exhibit behavior similar to the $\Lambda$CDM model.
\end{abstract}


\maketitle
\tableofcontents

\section{Introduction}

The cosmological standard model is currently based on the $\Lambda$CDM background, in which a (bare) cosmological constant, $\Lambda$, drives the cosmic speed up, and CDM stands for Cold Dark Matter \cite{SupernovaCosmologyProject:1998vns,SupernovaCosmologyProject:1997zqe,SupernovaSearchTeam:1998fmf,Peebles:2002gy}. The model appears particularly elegant, encompassing the majority of experimental tests \cite{Planck:2019nip,Planck:2019evm,Planck:2018nkj,Planck:2018vyg, Carloni:2024zpl}, despite recently puzzled by rough evidences that may favor an evolving dark energy term \cite{DESI:2024mwx, Zhao:2017cud, Muccino:2020gqt, Xu:2021xbt, Copeland:2006wr, Avsajanishvili:2023jcl,Dunsby:2016lkw,Colgain:2021pmf,Bamba:2012cp,Elizalde:2004mq,Wolf:2024eph}.

Accordingly, the revived interest in dark energy models, i.e., models in which dark energy is time-evolving may shed light on worrisome inconsistencies, such as cosmological tensions \cite{DiValentino:2021izs, Bernal:2016gxb, Vagnozzi:2019ezj,Heymans:2020gsg,KiDS:2020suj, Hildebrandt:2018yau, DiValentino:2017oaw}, physical interpretations of $\Lambda$ \cite{Sakharov:1967pk, Weinberg:2000yb, Carroll:2000fy, Dolgov:1997za, Sahni:1999gb, Straumann:1999ia, Rugh:2000ji,Padmanabhan:2002ji, Yokoyama:2003ii, Martin:2012bt}, existence of early dark energy \cite{Poulin:2018cxd,Karwal:2016vyq, Sohail:2024oki}, matching between dark energy and inflationary models \cite{Huterer:2002wf,Yokoyama:2002ts,Giare:2024akf,Brax:2024rqs}, and so forth, see e.g. \cite{Perivolaropoulos:2021jda, Bull:2015stt, Lynch:2024hzh, zabat:2024wof, Wang:2024dka, Allali:2024cji, Clifton:2024mdy, Carrilho:2023qhq, Lyu:2020lwm,Wolf:2023uno}.

In that matter, the preliminary release of the DESI collaboration has unexpectedly highlighted that a possible $w_0w_a$CDM model appears more compatible than the $\Lambda$CDM paradigm by considering new baryonic oscillation data catalogs \cite{DESI:2024mwx}, by reinterpreting dark energy in terms of an \emph{evolving scalar field} \cite{Copeland:2006wr}. However, it seems that the simplest evolving scalar field models do not alleviate the $H_{0}$ tension, as described in Ref. \cite{Lee:2022cyh}.

In this respect, although theoretically well-established, non-minimally coupled scalar field dark energy scenarios have not been extensively explored at late-time \cite{Szydlowski:2008in,Setare:2010pfa,Wolf:2024stt}, but most of the analyses have been pursued at early times\footnote{Non-minimal couplings introduce a further fifth force, allowing the interaction between curvature and scalar field \cite{Hrycyna:2015vvs}. Further, excluding Jordan or Einstein worsens the use of non-minimal couplings that may appear complicated to interpret \cite{Luongo:2024opv}.
} \cite{Kaiser:2015usz,Ema:2017loe,Bahamonde:2017ize,DiValentino:2019jae}.

Although not conclusive evidence for inflation, recent findings from the Planck satellite suggest that a non-minimal coupling between scalar fields and curvature significantly enhances the viability of chaotic potentials, which would otherwise be ruled out by observational data\cite{Panotopoulos:2014hwa, Nozariand:2015yfi, Eshaghi:2015rta, delCampo:2015wma, Bostan:2023ped, Kamali:2018ylz}. Remarkably, in the context of Higgs inflation, a fourth-order potential non-minimally coupled to the curvature leads to the Starobinsky potential, shifting to the Einstein frame \cite{Calmet:2016fsr, Mantziris:2022fuu,Mishra:2019ymr}. To put into perspective, non-minimal couplings may help in both alleviating the cosmological tensions, being quite important even at late-time \cite{DiValentino:2019jae, BarrosoVarela:2024htf} and not only immediately after the Big Bang.

Hence, embracing the idea of exploring possible couplings between dark energy fields and curvature, at both late and intermediate times, would therefore open new avenues toward the fundamental properties related to dark energy, among which its evolution \cite{Capozziello:2022jbw}, a possible interaction with other sectors, its form, and so on  \cite{Farrar:2003uw, Wang:2016lxa}.
Motivated by the above points, we here focus on scalar field dark energy models, highlighting the main differences when coupling dark energy with curvature, $R$, torsion, $T$, and non-metricity, $Q$, scalars.

We explore six quintessence scenarios and five phantom potentials of dark energy, whose evolution has not been ruled out by observations yet, but rather can revive as due to the last findings presented by the DESI collaboration.  For the sake of completeness, we additionally seek for dust-like scalar fields, namely for alternative version of scalar field description, that is divided into \emph{quasiquintessence} and \emph{quasiphantom field} \cite{Luongo:2023jnb}, respectively. Precisely, in both the so-cited representations, we analyze the corresponding modified Friedmann equations and study the cosmological dynamics by computing autonomous systems of first-order differential equations. Reliably, in formulating the dynamical variables, in lieu of using the widely-used exponential and power-law potentials \cite{Copeland:2006wr, Bahamonde:2017ize}, whose advantage only lies in reducing the overall complexity, we select more appropriate dimensionless functions that may work regardless the types of involved potentials. To this end, we explore the critical exponents and fix the free parameters for each potential. Our main findings suggest that, under minimal coupling, the quasiquintessence model reveals a critical point where the universe behaves like dust. This critical point does not appear in the standard scalar field approach, indicating that an alternative framework, in the form of quasiquintessence, could properly unify dark energy and dark matter under the same scenario. This dust-like characteristic  disappears when coupling is introduced, hiding the differences between the standard and alternative descriptions. Here, our results show that, while it is possible for dark energy to couple with curvature, torsion, or non-metricity scalars, not all the paradigms appear viable. Notably, in the phantom field scenario, numerical analysis yields inconclusive results for power-law potentials within the curvature-coupling framework. In contrast, only the inverse square potential is not supported in the teleparallel and symmetric-teleparallel dark energy.

The paper is structured as follows. In Sect.~\ref{Sec2}, we introduce the non-minimally couplings within the theories of gravity and analyze the modified cosmological equations. In Sect.~\ref{Sec3}, we study the modified cosmological equations and present the autonomous systems for both minimally and non-minimally coupled cosmologies. Last but not least, we also introduce all the potentials used for quintessence and phantom field. Then, in Sect.~\ref{Sec4}, we perform a stability analysis for the minimally and non-minimally coupled scenarios, identifying the critical points for all the potentials and computing the cosmological quantities within them.  In Sect.~\ref{Sec5}, we consider the growth of matter perturbations comparing our dark energy models with the concordance paradigm. Finally, in Sect.~\ref{Sec6}, we present the conclusions and perspectives of our work.


\section{Non-minimal couplings of dark energy models}\label{Sec2}

In this section, we report the different geometrical quantities to describe gravity, i.e., curvature, $R$, torsion, $T$, and non-metricity, $Q$, scalars. Moreover, we introduce below the non-minimal couplings between the scalar field, acting as dark energy, and gravity, under the form of $R, T$, and $Q$, properly chosen to describe each background distinctly. The non-minimal coupling between gravity and scalar fields is a subject of great speculation and study. Understanding \emph{a priori} how to make a non-minimal coupling up is a complicated task and, so, in analogy to quantum field theory, likely the simplest approach lies in considering a Yukawa-like interaction. The latter consists of a coupling between two fields, whose interaction is short-range, providing a non-dimensional coupling constant that, in principle, would guarantee that renormalization works. The original idea to characterize such an interaction deals with the coupling between fermions and bosons, whereas the analogy is based on substituting to one field the Ricci curvature and to the other the phion associated with dark energy. This procedure, widely adopted in the literature, see e.g. \cite{Luongo:2024opv, Belfiglio:2024swy}, and references therein, opens new avenues toward the frame problem, but presents probably the simplest way to model a non-minimal coupling, compatible with the recipe argued from field theories.

\subsection{Background}

At the background, action for non-minimally coupled theories of gravity is given by
\begin{align}
    &\mathcal{S}=\int d^{4}x \sqrt{-g}\left[\frac{R}{2k^{2}}-\frac{\xi R \phi^{2}}{2}+\mathcal{L}_{m}+\mathcal{L}_{\phi}\right],\\
    &\mathcal{S}=\int d^{4}x \sqrt{-g} \left[\frac{T}{2k^{2}}-\frac{\xi T \phi^{2}}{2}+\mathcal{L}_{m}+\mathcal{L}_{\phi}\right],\\
     &\mathcal{S}=\int d^{4}x \sqrt{-g}\left[\frac{Q}{2k^{2}}-\frac{\xi Q \phi^{2}}{2}+\mathcal{L}_{m}+\mathcal{L}_{\phi}\right].
\end{align}
Therefore, in our context, it would be convenient to set the general action, $\mathcal{S}$, depending on $R$, $T$ and $Q$, as
\begin{equation}\label{eq:genaction}
\mathcal{S}=\int d^{4}x \sqrt{-g}\left[\frac{\mathcal F(U,\phi)}{2k^{2}}+\mathcal{L}_{m}+\mathcal{L}_{\phi}\right],
\end{equation}
where $k^{2}=8\pi G$ and $\mathcal F(U,\phi)=U-k^{2}\xi U\phi^{2}$, whereas $U\equiv\{R, T, Q\}$.

This notation is used to represent extended theories of gravity; however, it is also useful for expressing the non-minimal coupling between the scalar field and gravity.

Indeed, Eq.\eqref{eq:genaction} represents an auxiliary functional, depending on the scalar field and on the ``gravity function'', identified in the set $U$. Clearly, $U$ singles out one  type of gravitational interaction, among the three and, moreover, in $\mathcal F$, the coupling constant, $\xi$, represents the strength of the particular fifth force between $\phi$ and $U$. Accordingly, we implicitly refer to the tenet that, for $R, T$ and $Q$, the interaction strength remains unaltered.

Further, $\mathcal{L}_{m}$ and $\mathcal{L}_{\phi}$ denote the matter and scalar field sectors, respectively, indicating the kind of non-gravitational contributions entering the energy-momentum tensor.

We below derive the corresponding cosmological equations for a non-minimally coupled scalar field, adopting the spatially flat version of the Friedmann-Robertson-Walker (FRW) metric,
\begin{equation}
    ds^2=dt^2-a(t)^2(dr^2+r^2d\Omega^2),\label{frw}
\end{equation}
where in the above-adopted radial coordinates, the angular part reads $d\Omega^2\equiv d\theta^2+\sin^2\theta d\phi^2$.

Under this scenario, we consider a dark energy field constructed by two different representations, i.e., the \emph{standard one} and an \emph{alternative view}, with the great advantage to mime dust, having that the sound speed identically vanishes\footnote{The importance of alternatives to standard cases is a recent development \cite{Farrar:2003uw, Gao:2009me, Wang:2016lxa, Dunsby:2016lkw, Luongo:2018lgy, DAgostino:2022fcx}. Essentially, it lies on guaranteeing structures to form at all scales, as certified by observations \cite{Daniel:2010ky}. Conversely to standard scalar field, behaving as stiff matter, a zero sound speed enables the Jeans length to be zero, making structures possible at wider scales, see Refs. \cite{Creminelli:2009mu, Camera:2012sf}.} \cite{Luongo:2018lgy, Luongo:2024opv, DAgostino:2022fcx, Luongo:2023jnb, Aviles:2011ak}.

Quite remarkably, we here observe that moving beyond general relativity into $\mathcal{F}(U,\phi)$ theories disrupts the equivalence described by the geometrical trinity of gravity. In other words, the non-minimal coupling acts to modify the backgrounds, breaking the equivalence among the three descriptions that, as well-known, are fully-equivalent in the minimal case. This may occur since the auxiliary function $\mathcal{F}(U,\phi)$ can be mapped onto the well-known $f(R)$, $f(T)$, and $f(Q)$ theories, where the scalar field dynamics acts to modify the background itself. Clearly, as $\xi \rightarrow 0$, the geometrical trinity of gravity is recovered, namely the physical equivalence among backgrounds is restored. The non-equivalence, however, has consequences on the stability, showing that the three theories deserve more investigation, taken case by case. This fact is also in conflict with the issue of passing from the Jordan to the Einstein frames, showing different physical descriptions for different theories in alternative frames, see e.g. \cite{Faraoni:1999hp, Belfiglio:2024swy, Capozziello:2019cav}. Limiting to the Jordan frame, we will focus on this problem later in the text, developing small perturbation analysis and checking the goodness of each model in different background scenarios.

In our work, we consider therefore three main cases, corresponding to three different background scenarios. Indeed, identifying gravity with the curvature suggests a naive interpretation of gravity as curvature spacetime, i.e., permits one to recognise the gravitational phenomena by having a curved spacetime, in fulfillment of the equivalence principle.

The latter implies that the particles move as due to geometrical properties imposed by postulating a given spacetime.

In this respect, invoking the equivalence principle, one can wonder whether
equivalent manners can be used to geometrize gravity since it is possible to recall
that a given spacetime is endowed with a metric and an affine structure, determined by
\begin{align}
    &g_{\mu\nu},\quad {\rm Metric}\notag\\
    &\Gamma^{\alpha}_{\mu\nu}\quad {\rm Connection}.
\end{align}

that turn out to be  completely independent between them.

For example, this point is evident while in extended theories of gravity one can argue different equations of motion in the Palatini formalism, i.e., by varying the action with respect to $\Gamma^\alpha_{\mu\nu}$, instead of $g_{\mu\nu}$.

In this respect, we can admit that if the connection \emph{is not} metric, then we account for non-metricity theories. On the other side, when the  antisymmetric part of the connection is present we define the torsion. Respectively, we have:
\begin{align}
Q_{\alpha\beta\gamma}=\nabla_\alpha g_{\beta\gamma},\\ T^{\alpha}_{\beta\gamma}=2\Gamma_{[\beta,\gamma]}^{\alpha}.
\end{align}

At this stage, one can consider then, among all possible connections that can be defined on a spacetime, the Levi–Civita connection. This appears the
unique connection both symmetric and metric-compatible. Hence, to describe gravity adopting a given metric, in our case using the maximally-symmetric FRW, it is possible to invoke the standard procedure permits to rewrite the torsion and non-metric theories using a given metric. For example, in the case of torsion, invoking the vierbein field, related to the spacetime metric. This enables one to use the FRW, without making the overall analysis made for stability unphysical. In other words, it is possible to choose a given spacetime and to use it even for $T$ and $Q$, as established in Refs. \cite{Xu:2012jf,Hrycyna:2008gk,Carloni:2023egi,DAgostino:2018ngy,Lu:2019hra,Ghosh:2023amt}, avoiding \emph{de facto} gauge choices on $\Gamma^\alpha_{\beta\gamma}$ that would imply unphysical results, see Ref. \cite{BeltranJimenez:2019esp}.

\subsection{Scalar field dark energy}

As above claimed, it would be interesting to work two forms of scalar fields out, in order to describe dark energy.

We hereafter summarize them by:
\begin{itemize}
\item[-] The Barrow representation, representing the \emph{standard} scalar field picture, characterized by a density, $\rho_{\phi}$, and a pressure, $p_{\phi}$, is given by \cite{Copeland:2006wr}
\begin{subequations}
\begin{align}
&\rho_{\phi} = \epsilon \frac{\dot{\phi}^2}{2} + V(\phi),\
&p_{\phi} = \epsilon \frac{\dot{\phi}^2}{2} - V(\phi), \label{eq:standardscalar}
\end{align}
\end{subequations}
where conventionally $\epsilon = \pm 1$ denoting quintessence and phantom field, respectively.

\item[-] The \emph{alternative} scalar field picture is given by \cite{Luongo:2018lgy, DAgostino:2022fcx}
\begin{subequations}
\begin{align}
&\rho_{\phi} = \epsilon \frac{\dot{\phi}^2}{2} + V(\phi),\
&p_{\phi} = -V(\phi), \label{eq:alterscalar}
\end{align}
\end{subequations}
where $\epsilon = \pm 1$ indicates as above.
\end{itemize}

Both the representations are included by displaying the following generic Lagrangian,

\begin{equation}\label{lagra2}
    \mathcal L = K-V(\phi)+\lambda Y[X,\nu(\phi)],
\end{equation}
where $X\equiv \epsilon\frac{\dot \phi^2}{2}$ and $K\equiv K(X,\phi)$, while $\lambda$ is a Lagrange multiplier, i.e., zero for the standard case and nonzero for the alternative one \cite{Gao:2009me, Luongo:2018lgy}. Here, $\nu(\phi)$ represents the chemical potential and $Y$ a generic functional of it.

Thus, selecting $K=X$, from Eq. \eqref{lagra2}, and $\lambda=0$, for the standard case, we end up with
\begin{equation}\label{lagranew1}
    \mathcal L = \epsilon\frac{\dot \phi^2}{2}-V(\phi),
\end{equation}
while, restoring $\lambda$, we obtain
\begin{equation}\label{lagranew2}
    \mathcal L = \epsilon\frac{\dot \phi^2}{2}-V(\phi)+\lambda Y,
\end{equation}
that generically refers to the alternative description, yielding a quasiquintessence fluid when $\epsilon=+1$ and a quasiphantom fluid for $\epsilon=-1$.

To comprehend the idea behind the names quasiquintessence and quasiphantom, one can investigate the corresponding properties. In particular, the last form in Eq.~\eqref{lagranew2}, with $\epsilon=+1$, can be arguable either from modifying the Einstein equations by hand, see Ref. \cite{Gao:2009me}, or by invoking an energy constraint imposed by virtue of $Y$ \cite{Luongo:2018lgy}. The  scenario yields a \emph{dust-like behavior} of dark energy, in fact, from Eqs. \eqref{lagranew1} and \eqref{lagranew2}, computing the sound speed, $c_s^2\equiv\frac{\partial p}{\partial \rho}$, we immediately find,
\begin{subequations}
    \begin{align}
        c_{s,Q}^2&= \frac{\partial p_\phi }{\partial X}/\left(\frac{\partial \rho_\phi }{\partial X}\right)=1,\\
        c_{s,QQ}^2&= \frac{\partial p_\phi }{\partial X}/\left(\frac{\partial \rho_\phi }{\partial X}\right)=0,
    \end{align}
\end{subequations}
respectively for quintessence, $c_{s,Q}$ and quasiquintessence, $c_{s,QQ}$. Interestingly, the same can be generalized for the quasiphantom case, by simply invoking  $\epsilon=-1$ since the very beginning.

In the quintessence case, then, dark energy behaves as stiff matter, as the sound speed is unitary, i.e., resembles the speed of light, whereas in the quasiquintessence or quasiphantom cases, dark energy acts as a matter-like fluid, since the sound speed identically vanishes, as for dust\footnote{These properties show severe implications in matter creation throughout the universe evolution, see Refs. \cite{Belfiglio:2022qai, Belfiglio:2023rxb}.} \cite{DAgostino:2022fcx, Luongo:2024opv}.

In addition, under particular circumstances, the quasiquintessence naturally arises to heal the cosmological constant problem, as shown in Refs. \cite{Luongo:2018lgy,Belfiglio:2023rxb}, while being able to drive inflation as well,  throughout the existence of a metastable phase, see Refs. \cite{DAgostino:2022fcx,Luongo:2024opv}.

\subsection{Non-minimally coupled cosmology in Einstein's theory}

In general relativity, gravity is described with the scalar curvature, so we select $U=R$.

Considering here the non-minimal coupling scenario \cite{Nojiri:2006ri, Sotiriou:2008rp, DeFelice:2010aj, Szydlowski:2008in, Setare:2010pfa,Nojiri:2010wj,Nojiri:2017ncd}, the field equations obtained by varying the modified Hilbert-Einstein action in Eq.~\eqref{eq:genaction} with respect to a generic spacetime read
\begin{widetext}
\begin{equation}
G_{\mu\nu}\mathcal F+\left(\partial_{\mu}\partial_{\nu}-g_{\mu\nu}\partial_{\mu}\partial^{\mu}\right)\mathcal F_{R}-\frac{1}{2}g_{\mu\nu}\left[\mathcal F-R \mathcal F_{R}\right]=k^{2}\left(T^{(m)}_{\mu\nu}+T^{(\phi)}_{\mu\nu}\right).
\end{equation}
\end{widetext}

Here, $G_{\mu\nu}$ is the Einstein tensor, the subscript $R$ denote the derivatives with respect to the scalar curvature and $T^{(m)}_{\mu\nu}$ and $T^{(\phi)}_{\mu\nu}$ represent the energy-momentum tensor for matter and scalar field, respectively.

At this stage, assuming Eq. \eqref{frw}, we obtain the modified Friedmann equations, i.e.,
\begin{align}
        &H^{2}=\frac{1}{3 \mathcal F_{R}}\left(k^2\rho_t+\frac{R \mathcal F_{R}-\mathcal F}{2}-3H{\dot{\mathcal F}_{R}}\right),\label{eq:FRW10}\\
        &2\dot{H}+3H^{2}=-\frac{k^{2}p_t+\ddot{\mathcal F}_{R}+2H\dot{\mathcal F}_{R}+(1/2)(\mathcal F-R \mathcal F_{R})}{\mathcal F_{R}}, \label{eq:FRW20}
\end{align}
in which $H=\frac{\dot{a}}{a}$ is the Hubble parameter, $\rho_t$ and $p_t$ are the total density and pressure, respectively given by: $\rho_t\equiv \rho_{\phi}+\rho_{m}$ and $p_{\phi}+p_{m}$, whereas the dot indicates the derivative with respect to the cosmic time $t$.

By substituting the explicit form of $\mathcal F\left(R,\phi\right)$ and recalling moreover that $R=6\left(\dot{H}+2H^{2}\right)$ in a homogeneous and isotropic universe \cite{Geng:2011aj},  Eqs.~\eqref{eq:FRW10} and~\eqref{eq:FRW20} can be rewritten as
\begin{align}
      & H^{2}=\frac{k^{2}}{3}\left(\rho^{\rm eff}_{\phi}+\rho_{m}\right),\label{eq:FRW11}\\
        &2\dot{H}+3H^{2}=-k^{2}\left(p^{\rm eff}_{\phi}+p_{m}\right),\label{eq:FRW21}
\end{align}
yielding the net effective density and pressure,
\begin{align}
    &\rho^{\rm eff}_{\phi}=\rho_{\phi}+6\xi H\phi \dot{\phi}+3\xi H^{2}\phi^{2},\label{effrho}\\
    &p^{\rm eff}_{\phi}=p_{\phi}-2\xi \phi \ddot{\phi}-2\xi \dot{\phi}^{2}-4H\xi \phi\dot{\phi}-\xi\left(2\dot{H}+3H^{2}\right)\phi^{2},\label{eq:effpre}
\end{align}
provided  $\rho_{\phi}$ and $p_{\phi}$ from Eqs. \eqref{eq:standardscalar} and  \eqref{eq:alterscalar}, for both $\epsilon=\pm1$.

Hence, by varying Eq.~\eqref{eq:genaction} with respect to the scalar field, we find two  modified Klein-Gordon equations, both exhibiting a source term,
  \begin{equation}
\ddot{\phi}+\frac{3}{\sigma}H\dot{\phi}+\epsilon V_{,\phi}=-6\epsilon\xi\left(\dot{H}+2H^{2}\right)\phi,\label{eq:modKG}
\end{equation}
in which $\sigma=1$ and $\sigma=2$ select the kind of field, namely are given for Eqs.~\eqref{eq:standardscalar} or Eq.~\eqref{eq:alterscalar}, respectively.

Remarkably, combining Eqs.~\eqref{eq:FRW10} and~\eqref{eq:FRW20} provides the Raychaudhuri equation,
\begin{widetext}
\begin{equation}
    \dot{H}=-\frac{k^{2}}{1+\xi\left(6\xi-1\right)\left(k\phi\right)^{2}}\left[\left(\epsilon-2\sigma\xi\right)\frac{\dot{\phi}^{2}}{2\sigma}+4\xi H \phi\dot{\phi}+12\xi^{2} H^{2}\phi^{2}+\xi \phi \epsilon V_{,\phi}+\frac{(1+w_{m})}{2}\rho_{m}\right],
\end{equation}
\end{widetext}
where $w_{m}$ is conventionally the barotropic factor for matter\footnote{For the sake of completeness, the matter sector is not specified yet at this stage. It appears clear that dust provides $w_m=0$, albeit other kinds of matter contribution may furnish different barotropic factor. In general, the term $w_m$ may also indicate a barotropic fluid satisfying the Zeldovich conditions on the equation of state.}.

\subsection{Non-minimally teleparallel and symmetric-teleparellel cosmology}

Compactly the non-minimally teleparallel and symmetric-teleparellel cases can be treated in the same way, from the mathematical viewpoint, once the coincidence gauge is optioned for the latter treatment  \cite{BeltranJimenez:2017tkd}. Even though aware of the conceptual difficulties related to the coincident gauge stability \cite{Heisenberg:2023wgk}, we focus on this case only in order to show the formal degeneracy between $T$ and $Q$ in framing the non-minimal coupling.

Below we treat in detail both the cases.

\paragraph*{The teleparallel case.}

In teleparallel gravity \cite{Bahamonde:2021gfp, Clifton:2011jh, Paliathanasis:2021nqa, Hohmann:2018rwf, Hohmann:2019nat,Krssak:2015oua},
the metric $g_{\mu\nu}$ is explicitly expressed in terms of the vierbein field $e^A_\mu$ by $g_{\mu\nu} =  e^A_\mu e^B_\nu\eta_{AB}$,
where $\eta_{AB} = \text{diag} (1, -1, -1, -1)$, and the determinant of the vierbein is $e = \det(e^A_\mu) = \sqrt{-g}$.

By introducing the spin connection $\omega^A_{\ B\mu}$, which governs the rule of parallel transportation, we can determine the torsion tensor and the corresponding contorsion tensor as follows
\begin{align}
    &T^A_{\ \mu\nu}(e^A_{\ \mu},\omega^A_{\ B\mu})=
\partial_\mu e^A_{\ \nu} -\partial_\nu e^A_{\ \mu}+\omega^A_{\ B\mu}e^B_{\ \nu}
-\omega^A_{\ B\nu}e^B_{\ \mu},\\
    &K^{\mu\nu}_{\  \ A}=\frac{1}{2}
\left(T^{\ \mu\nu}_{A}
+T^{\nu\mu}_{\ \ A}
-T^{\mu\nu}_{\  \ A}\right).
\end{align}

Here, the Lagrangian density is characterized by the torsion scalar $T$, in contrast to $R$ used in general relativity. The torsion scalar
$T$ is thus given by
$T= T^A_{\ \mu\nu}S_A^{\ \mu\nu}$, in which
    $S_A^{\ \mu\nu}= e_A^{\ \rho}S_\rho^{\
\mu\nu}=K^{\mu\nu}_{\ \ A}
-e^A_{\ \nu}T^{ \mu}
+e^A_{\ \mu}T^{ \nu}$. Now, the vierbein field $e^A_\mu = (1, -a^2, -a^2, -a^2)$, which is in the Weitzenböck gauge with a vanishing spin connection, is the only choice that ensures both the vierbein field and the teleparallel connection respect the cosmological symmetries for a flat FRW metric \cite{Bahamonde:2021gfp}. Then, by varying  Eq.~\eqref{eq:genaction} with respect to the tetradal $e^A_\nu$, we end up with the modified Einstein's equations
\begin{widetext}
\begin{equation}
e^{-1}\partial_{\mu}\left(e e^\rho_A S_{\rho}{}^{\mu\nu}\right)\mathcal F_{T}- e_{A}^{\lambda}T^{\rho}{}_
{\mu\lambda}S_{\rho}{}^{\nu\mu}\mathcal F_{T}
        + e^\rho_A S_{\rho}{}^{\mu\nu}(\partial_{\mu}{T})\mathcal F_{TT}+\frac{1}{4}e_{A}^{\nu}\mathcal F=\frac{k^{2}}{2} e_{A}^{\rho}{\left(T^{(m)}+T^{(\phi)}\right)}_{\rho}{}^{\nu},
\end{equation}
\end{widetext}
where the subscript $T$ represents the derivatives with respect to the torsion scalar, and the modified Friedmann equations become \cite{Mirza:2017afs, Geng:2011aj,Gu:2012ww,Leon:2022oyy},
\begin{align}
    &6H^{2}\mathcal F_{T}+\frac{1}{2}\mathcal F=k^{2}\rho_{m},\\
    &2{\mathcal F}_{T}\left(3H^{2}-\dot{H}\right)+2H\dot{\mathcal F}_{T}+\frac{1}{2}\mathcal F=-k^{2}p_{m}.
\end{align}

Therefore, by substituting the explicit form of $\mathcal F$ and recalling that $T=-6H^{2}$ in the flat FRW metric, we can rewrite the above equations as
\begin{align}
    &H^{2} = \frac{k^{2}}{3}\left(\rho^{\rm eff}_{\phi} + \rho_{m}\right)\label{eq:FRW11T},\\
    &2\dot{H} + 3H^{2} = -k^{2}\left(p^{\rm eff}_{\phi} + p_{m}\right)\label{eq:FRW21T},
\end{align}
where the effective density and pressure for the scalar field are given by
\begin{align}
     &\rho^{\rm eff}_{\phi} = \rho_{\phi} + 3H^{2}\xi \phi^{2},\\
    &p^{\rm eff}_{\phi} = p_{\phi} - 4H\xi \phi \dot{\phi} - \left(3H^{2} + 2\dot{H}\right)\xi \phi^{2}.
\end{align}

Furthermore, by varying Eq.~\eqref{eq:genaction} with respect to the scalar field $\phi$, we get the modified Klein-Gordon equation
\begin{equation}
    \ddot{\phi}+\frac{3}{\sigma}H\dot{\phi}+\epsilon V_{,\phi}=\epsilon 6 H^{2}\xi \phi,\label{eq:KGT}
\end{equation}
and by combining Eqs.~\eqref{eq:FRW11T} and~\eqref{eq:FRW21T}, we obtain the explicit  Raychaudhuri equation
\begin{equation}
    \dot{H}=-\frac{k^{2}\left(\epsilon\frac{\dot{\phi}^{2}}{\sigma}-4H\xi \phi \dot{\phi}+\left(1+w_{m}\right)\rho_{m}\right)}{2\left(1-\xi k^{2}\phi^{2}\right)}.
\end{equation}

\paragraph*{The symmetric-teleparallel case.}

All these equations are \emph{in form} even valid for the non-minimally symmetric-teleparallel coupled cosmology, in the coincident gauge\cite{Carloni:2023egi, Ghosh:2023amt, BeltranJimenez:2017tkd}, as stated in the beginning of this subsection. Particularly, in symmetric-teleparellel theory of gravity the connection is based only on the non-metricity, without any torsion \cite{Lu:2019hra, Koussour:2023hgl,Paliathanasis:2023nkb,Dimakis:2023cam,Millano:2024rog,Paliathanasis:2024xpy,Jarv:2018bgs}. This implies that the connection is symmetric, and the torsion tensor vanishes, as well as the Riemann tensor. The non-metricity tensor, given by $Q_{\rho\mu\nu}=\nabla_{\rho} g_{\mu\nu}$, is used to write the disformation
\begin{equation}
    L_{\rho\mu\nu}=\frac{1}{2}\left(Q_{\rho\mu\nu}-Q_{\mu\rho\nu}-Q_{\nu\rho\mu}\right),
\end{equation}
a part of the general affine connection.

Here, we can derive the invariant, i.e., the non-metricity scalar,
\begin{equation}
    Q=\frac{1}{2}P^{\rho\mu\nu}Q_{\rho\mu\nu},
\end{equation}
where $P^{\rho\mu\nu}$ is called superpotential, provided by
\begin{equation}
    P^{\rho\mu\nu}=L^{\rho\mu\nu}+\frac{1}{2}g^{\mu\nu}\left(\tilde{Q}^{\rho}-Q^{\rho}\right)+\frac{1}{4}\left(g^{\rho\mu}Q^{\nu}+g^{\rho\nu}Q^{\mu}\right),
\end{equation}
with $\tilde{Q}_{\rho}\equiv \tensor{Q}{^{\mu}_{\mu\rho}}$ and $Q_{\rho}\equiv \tensor{Q}{_{\rho}^{\mu}_{\mu}}$, dubbed  the non-metricity traces.

At this point, varying the action in Eq.~\eqref{eq:genaction} with respect to the metric yields
\begin{widetext}
\begin{equation}
    \frac{2}{\sqrt{-g}}\nabla_{\rho}\left(\sqrt{-g}\mathcal{F}_{Q} \tensor{P}{^{\rho}_{\mu\nu}}\right)
    + \frac{1}{2} g_{\mu\nu} \mathcal{F}
    + \mathcal{F}_{Q} \left(P_{\mu\rho\alpha} \tensor{Q}{_{\nu}^{\rho\alpha}}
    - 2Q_{\rho\alpha\mu} \tensor{P}{^{\rho\alpha}_{\nu}}\right)
    = k^{2} \left(T^{(m)}_{\mu\nu} + T^{(\phi)}_{\mu\nu}\right),
\end{equation}
\end{widetext}
where the subscript $Q$ denotes derivatives with respect to the non-metricity scalar.

In addition, varying Eq.~\eqref{eq:genaction} with respect to the symmetric flat connection gives the equation of motion for the connection,
\begin{equation}
    \nabla_{\mu} \nabla_{\nu} \left(\sqrt{-g} \mathcal{F}_{Q} \tensor{P}{^{\mu\nu}_{\rho}}\right) = 0,
\end{equation}
which is identically satisfied in the coincident gauge \cite{Paliathanasis:2023pqp}.

In this gauge, for FRW metric, the scalar takes the simple form $Q = -6H^{2}$, which formally resembles the torsion case. However, it is ``easier'' than general relativity due to the absence of the dynamical term $\sim \dot{H}$.

Consequently, we recover the same modified Friedmann equations along with the Klein-Gordon equation as in the non-minimally teleparallel framework—namely, Eqs.~\eqref{eq:FRW11T},\eqref{eq:FRW21T}, and \eqref{eq:KGT}—as observed in Refs. \cite{Carloni:2023egi,Ghosh:2023amt, BeltranJimenez:2017tkd, Paliathanasis:2023nkb}.


\section{Autonomous systems of dark energy models}\label{Sec3}

In this section, we present the autonomous systems for minimally and non-minimally coupled dark energy models. Specifically, the minimal case is considered with the alternative scalar field only, as it has not yet been studied. Additionally, for the non-minimally coupled framework, we focus on potentials that have not been extensively analyzed in the literature, highlighting the differences between the standard and alternative scalar field scenarios. The autonomous systems are described by first-order differential equations without explicit time-dependent terms.

Hence, the following path is worked out.

\begin{itemize}
\item[-] First, we consider the generic scenario of minimally coupled cosmology involving an alternative scalar field, which includes quasiquintessence and quasiphantom models. To account for all possible potentials, we employ a general analytical framework\footnote{The dynamics of the standard scalar field minimally coupled to gravity have been extensively studied in the literature, see, e.g., \cite{Copeland:2006wr,Bahamonde:2017ize}. However, investigations of the alternative scalar field have largely been limited to the quasiquintessence case, often utilizing the exponential potential, as discussed in \cite{Gao:2009me}.}. In this context, minimal coupling emerges as the limiting case of a non-minimally coupled framework where $\xi \ll 1$, recovering the standard Friedmann equations. Following this, we compare the results for the alternative scalar field with those derived for the standard minimally coupled scalar field, emphasizing the distinctions between the two approaches. Additionally, we expand our analysis to encompass all viable dark energy potentials that remain consistent with the most recent observational constraints on dark energy evolution.
\item[-] Second, we examine systems with non-minimal coupling to gravity, focusing on their effects on stability analysis. Specifically, we introduce a Yukawa-like coupling, assuming its applicability extends beyond general relativity to frameworks such as teleparallel and symmetric-teleparallel gravity, see Ref. \cite{Capozziello:2024mxh}.
\end{itemize}

The potentials, considered below, being not ruled out from observations, are thus used in order to explain late-time dark energy phenomenology. We include also phantom fields in our analysis, as they remain consistent with the recent DESI 2024 results. This is due to the fact that the $(w_0, w_a)$ parameters derived from the data provide no direct insight into the microphysics of the dark energy model. Distinguishing whether the model is phantom or not requires extending the best-fit $w(a)$ well beyond the redshift range used for parameter determination, as reported in \cite{Wolf:2024eph}.

\subsection{Quintessence-like potentials of dark energy}

Regarding quintessence, the potentials studied, which are shown in Fig. \ref{fig:Quint}, are given below.

\begin{itemize}
    \item[-] \emph{\underline{Sugra potential}}: The simplest positive scalar potential model, derived from supergravity, is known as the Sugra potential and has the form
    \begin{equation}  V\left(\phi\right)=\frac{\Lambda^{4+\chi}}{\phi^{\chi}}e^{\gamma k^{2} \phi^{2}},
    \end{equation}
    with $\chi, \gamma \gt 0$ . This potential form is often used to address the cosmological constant problem \cite{Martin:2012bt}. In the context of quintessence, the equation of motion with this potential leads to tracker solutions \cite{Brax:1999yv, Brax:1999gp, Caldwell:2005tm}.
    \item[-] \emph{\underline{Barreiro-Copeland-Nunes (BCN) potential}}: This is a double exponential potential defined as
\begin{equation}
V(\phi) = \Lambda^{4} \left( e^{l k\phi} + e^{m k\phi} \right),
\end{equation}
where $l, m \gt 0$ . In particular, a constraint arising
from nucleosynthesis requires that $l\gt 5.5$,  while $m \lt 0.8$ is needed to obtain a barotropic factor $w \lt -0.8$ in the quintessence scenario \cite{Barreiro:1999zs}.

    \item[-] \emph{\underline{Albrecht-Skordis (AS) potential}}: The AS potential takes the form
    \begin{equation}
    V\left(\phi\right)=\Lambda^{4}\left(\left(k\phi-B\right)^{2}+A\right)e^{-\mu k\phi},
    \end{equation}
    where $A\geq0$, $B\geq0$ and $\mu \gt0$. It belongs to a class of scalar field models that may naturally arise from superstring theory, explaining the acceleration observed in the present epoch \cite{Albrecht:1999rm}.  Additionally, we include the case where $A=0$ and $B=0$, as described in supergravity models, see e.g. \cite{Ng:2001hs}.
    \item[-] \emph{\underline{Ureña-López-Matos (ULM) potential}}: The quintessence potential given by
    \begin{equation} V\left(\phi\right)=\Lambda^{4}\left(\sinh^{n}\left(\zeta k\phi\right)\right),
    \end{equation}
    where $\zeta \gt0$ and $n\lt0$, is denoted as ULM potential.This potential acts as a tracker solution \cite{Urena-Lopez:2000ewq}, which may have driven the universe into its current inflationary phase. The ULM potential exhibits asymptotic behavior, resembling an inverse power-law potential in the early universe and transitioning to an exponential potential at late times.
     \item[-] \emph{\underline{Inverse exponential (IE) potential}}: The IE potential assumes the form
     \begin{equation}
     V\left(\phi\right)=\Lambda^{4}e^{\frac{1}{k\phi}},
     \end{equation}
    and it appears in Ref. \cite{Caldwell:2005tm}, where it is proposed as a tracker model, without a zero minimum.
      \item[-] \emph{\underline{Chang-Scherrer (CS) potential}}:  The CS potential is a modified exponential potential assuming the form
      \begin{equation}
V\left(\phi\right)=\Lambda^{4}\left(1+e^{-\tau k\phi}\right),
      \end{equation}
      where $\tau\gt0$. In contrast to models with a standard exponential potential, this approach even captures the features of early dark energy and may help to alleviate the coincidence problem, as shown in Ref. \cite{Chang:2016aex}.
\end{itemize}

\begin{figure*}
\centering
\subfigure[Sugra potential with $\chi = 1$ and $\gamma = 1$, and BCN potential with $m = 6$ and $l = \frac{1}{2}$.\label{Quint1}]
{\includegraphics[height=0.32\hsize,clip]{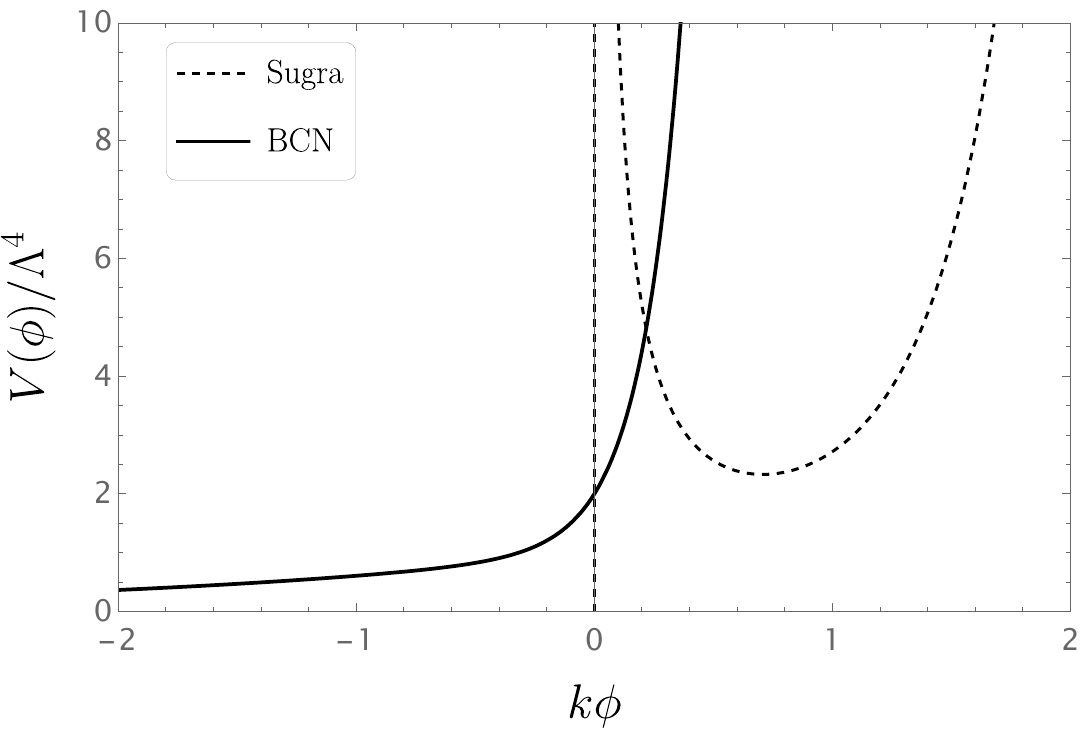}}
\hspace{2mm}
\subfigure[AS potential with $A = 0$, $B = 0$, and $\mu = 1$; ULM potential with $n = -\frac{1}{20}$ and $\zeta = 1$; IE potential; and CS potential with $\tau = 1$.\label{Quint2}]
{\includegraphics[height=0.32\hsize,clip]{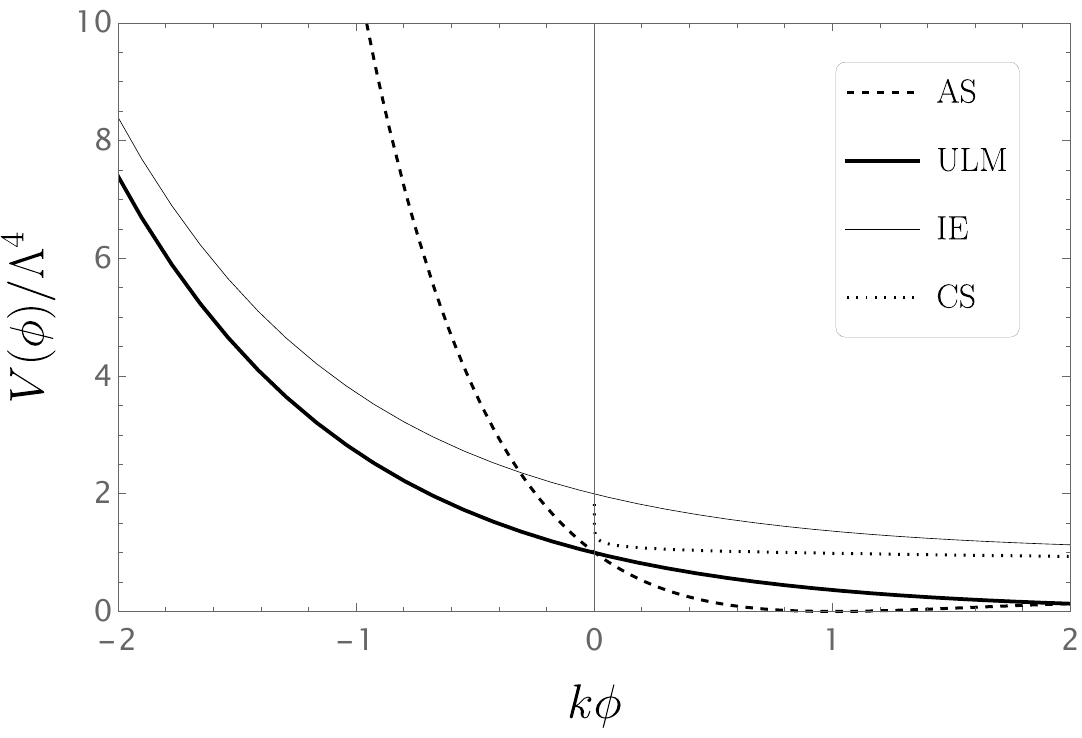}}
\caption{Behavior of quintessence potentials with fixed free parameters. The Sugra potential makes a significant contribution at both small and large field values, while the BCN potential is dominant only at large field values. For the Sugra potential, the offset is chosen as the Planck mass. All other potentials exhibit the same decreasing behavior as the field value increases.}\label{fig:Quint}
\end{figure*}

\subsection{Phantom-like potentials of dark energy}

\begin{figure*}
\centering
\subfigure[Power law potentials with $\alpha=5, 2$ and the exponential potential with $\beta=1$.\label{Phant1}]
{\includegraphics[height=0.32\hsize,clip]{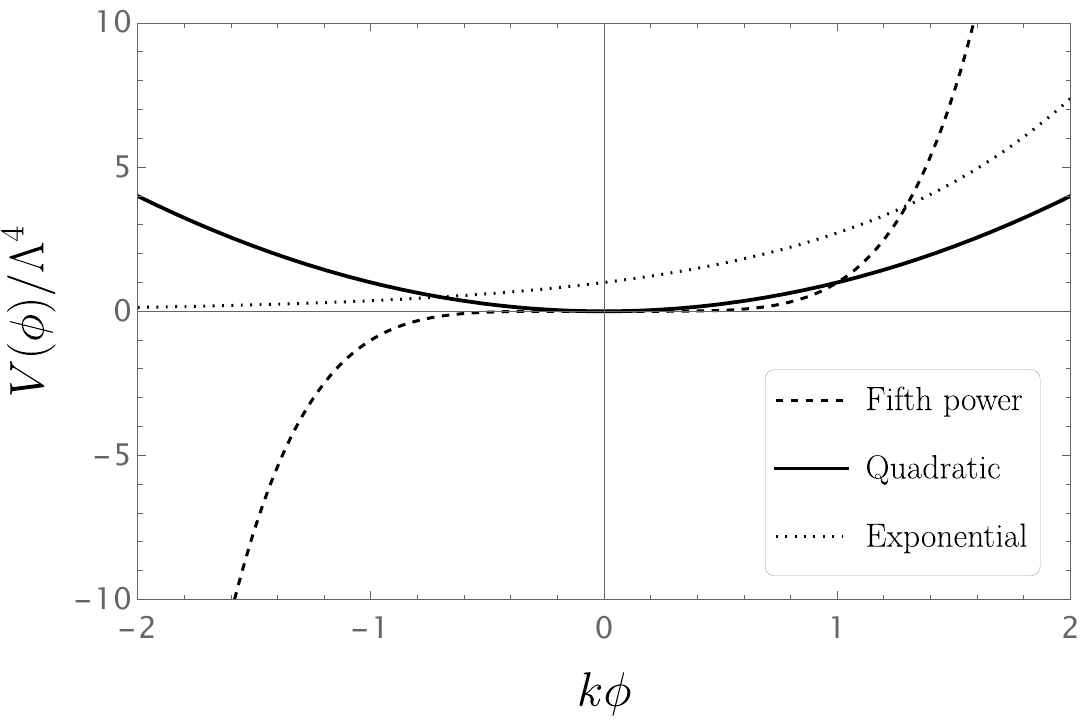}}
\hspace{2mm}
\subfigure[Power law potential with $\alpha=-2$, i.e., ISP potential, and the IHC potential with $\alpha=1$.\label{Phant2}]
{\includegraphics[height=0.32\hsize,clip]{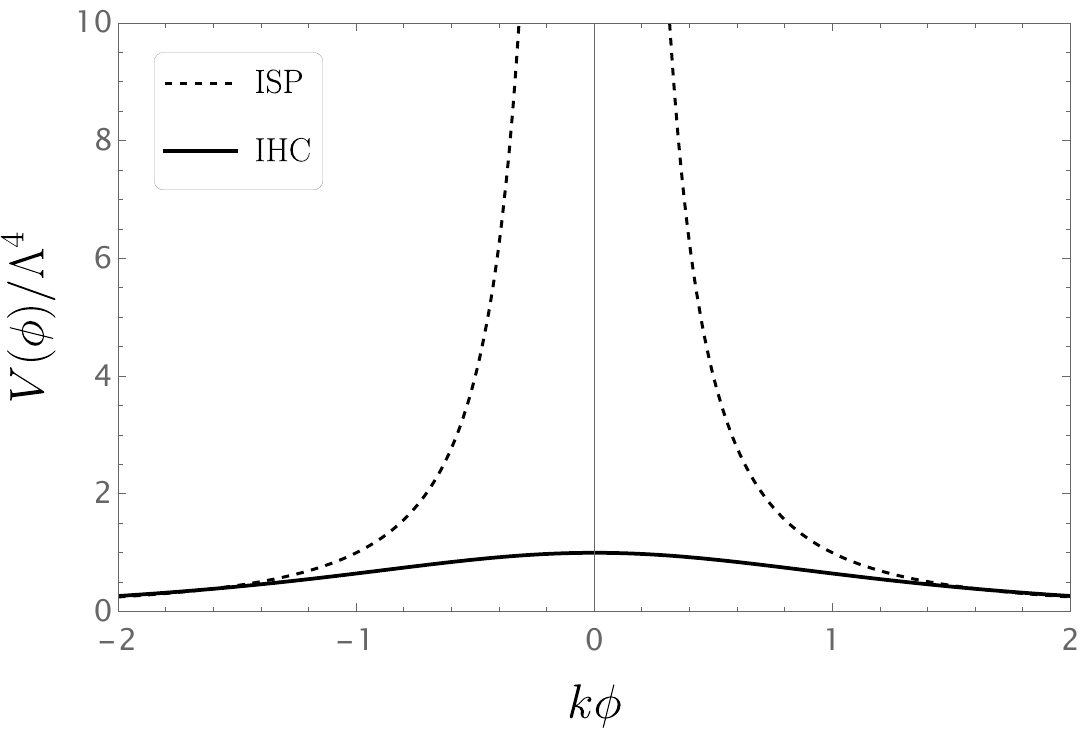}}
\caption{Behavior of phantom field potentials with fixed free parameters. The potentials that increase as the field grows are shown on the left, while those that decrease as the field increases are depicted on the right. The fifth-power potential is the only one that allows negative values.}\label{fig:Phant}
\end{figure*}

Conversely to quintessence-like behavior, for phantom field, we employ the following potentials, which are represented in Fig. \ref{fig:Phant}.

\begin{itemize}
    \item[-] \emph{\underline{Power law potentials}}: The expression for these types of potentials is given by
    \begin{equation}
V\left(\phi\right)=\Lambda^{4}\left(k\phi\right)^{\alpha},
    \end{equation}
and these potentials are well-fitted by the $w_{0}w_{a}$ parametrization at redshifts $z \lesssim 1$, as reported in Ref. \cite{Scherrer:2008be, Sami:2003xv}. Power-law potentials are divided into two classes: those with $\alpha \geq 4$, which lead to 'Big Rip' singularity, and those with $\alpha < 4$, which avoid it. In our work, we study $\alpha= 5, 2 , -2$, since they are not ruled out by observations.
    \item[-] \emph{\underline{Exponential potential}}: It is the usual exponential potential, i.e.,
    \begin{equation}
    V\left(\phi\right)=\Lambda^{4}e^{\beta k\phi},
    \end{equation} with $\beta \gt0$.This model, as for the
power law potentials, is well approximates by the
$w_{0}w_{a}$ parametrization, and it leads to a 'Big Rip' singluarity \cite{Scherrer:2008be,Sami:2003xv}.
     \item[-] \emph{\underline{Inverse hyperbolic cosine (IHC) potential}}: The inverse hyperbolic cosine potential takes the following form
     \begin{equation}
     V\left(\phi\right)=\Lambda^{4}\left(\cosh^{-1}\left(\alpha k\phi\right)\right),
     \end{equation} with $\alpha\gt0$. Here, the phantom field, when released from a point away from the origin with no initial kinetic energy, gravitates towards the peak of the potential, crosses it, and then reverses direction to undergo a damped oscillation around the potential's maximum \cite{Singh:2003vx}.

\end{itemize}

 \begin{table*}
    \centering
      \resizebox{0.9\textwidth}{!}{
\footnotesize
\setlength{\tabcolsep}{0.5em}
\renewcommand{\arraystretch}{2}
\begin{tabular}{|c|c|c|c|}
\hline
\hline
Potential & Expression & $\lambda$ & $\Gamma-1$\\
\hline\hline
\multicolumn{4}{|c|}{Type: \textit{Quintessence/Quasiquintessence}}\\
\cline{1-4}
\hline\hline
   Sugra \cite{Avsajanishvili:2023jcl, Brax:1999yv, Brax:1999gp, Caldwell:2005tm} &  $V\left(\phi\right)=\frac{\Lambda^{4+\chi}}{\phi^{\chi}}e^{\gamma k^{2} \phi^{2}}$ & $-2\gamma k\phi +\frac{\chi}{k \phi}$ & $\frac{4\gamma}{\lambda^{2}}+\frac{1}{\lambda k\phi}$\\
\hline
   BCN \cite{Avsajanishvili:2023jcl, Barreiro:1999zs} & $V\left(\phi\right)=\Lambda^{4}\left(e^{l k\phi}+e^{m k\phi}\right)$ & $-\frac{l e^{l k\phi}+m e^{m k\phi}}{e^{l k\phi}+e^{m k\phi}}$ & $-\frac{(l+\lambda)(m+\lambda)}{\lambda^{2}}$ \\
\hline
AS \cite{Avsajanishvili:2023jcl, Albrecht:1999rm} & $V\left(\phi\right)=\Lambda^{4}\left(\left(k\phi-B\right)^{2}+A\right)e^{-\mu k\phi}$ & $\mu + \frac{2\left(B-k\phi\right)}{A+\left(B-k\phi\right)^{2}}$ & $\frac{(\lambda -\mu ) \left(\frac{1}{B-k\phi }-\lambda +\mu \right)}{\lambda ^2}$ \\
\hline
ULM \cite{Avsajanishvili:2023jcl, Urena-Lopez:2000ewq} &
$V\left(\phi\right)=\Lambda^{4}\left(\sinh^{n}\left(\zeta k\phi\right)\right)$ & $-n \zeta \coth\left(\zeta k\phi\right)$ & $-\frac{1}{n}+\frac{n \zeta^{2}}{\lambda^{2}}$ \\
\hline
IE \cite{Avsajanishvili:2023jcl, Caldwell:2005tm}&
$V\left(\phi\right)=\Lambda^{4}e^{\frac{1}{k\phi}}$ & $\frac{1}{\left(k\phi\right)^{2}}$ & $\frac{2}{\sqrt{\lambda}}$ \\
\hline
CS \cite{Avsajanishvili:2023jcl, Chang:2016aex} & $V\left(\phi\right)=\Lambda^{4}\left(e^{-\tau k\phi}+1\right)$ & $\frac{\tau}{1+e^{\tau k \phi}}$ & $\frac{\tau}{\lambda}-1$ \\
\hline\hline
\multicolumn{4}{|c|}{Type: \textit{Phantom/Quasiphantom field}}\\
\hline\hline
Fifth power \cite{Avsajanishvili:2023jcl,Scherrer:2008be} & $V\left(\phi\right)=\Lambda^{4}\left(k\phi\right)^{5}$ & $-\frac{5}{k\phi}$ & $-\frac{1}{5}$ \\
\hline
ISP \cite{Avsajanishvili:2023jcl,Scherrer:2008be} & $V\left(\phi\right)=\Lambda^{4}\left(k\phi\right)^{-2}$ & $\frac{2}{k\phi}$ & $\frac{1}{2}$ \\
\hline
Exponential \cite{Avsajanishvili:2023jcl,Scherrer:2008be} & $V\left(\phi\right)=\Lambda^{4}e^{\beta k\phi}$ & $-\beta$ & $0$ \\
\hline
Quadratic \cite{Avsajanishvili:2023jcl,Scherrer:2008be} & $V\left(\phi\right)=\Lambda^{4}\left(k\phi\right)^{2}$ & $-\frac{2}{k\phi}$ & $-\frac{1}{2}$ \\
\hline
IHC \cite{Avsajanishvili:2023jcl,Singh:2003vx} &
$V\left(\phi\right)=\Lambda^{4}\left(\cosh^{-1}\left(\alpha k\phi\right)\right)$ & $\alpha \tanh\left(\alpha k \phi\right)$ & $1-\frac{\alpha^{2}}{\lambda^{2}}$ \\
\hline\hline

\end{tabular}}
\caption{Quintessence and phantom field potentials in minimally and non-minimally coupled cosmologies.}\label{tab:potchar}
\end{table*}

\subsection{Minimally coupled cosmology}

We analyze the dynamics of scalar field by rewriting the Klein-Gordon equation in terms of a set of dimensionless variables that, in general, are arbitrarily chosen.

In particular, to determine the dynamics of the scalar field, we might solve an autonomous system composed of first-order differential equations, involving dimensionless variables evolution.

In this subsection, we analyze the alternative scalar field description minimally coupled to gravity. To do so, we consider the dimensionless variables given by \cite{Gao:2009me, Copeland:2006wr, Bahamonde:2017ize, Boehmer:2014vea}
\begin{align}
&x\equiv \dfrac{k\dot{\phi}}{\sqrt{6}H}\,, \hspace{1cm} y\equiv \dfrac{k\sqrt{V}}{\sqrt{3}H}\,,\hspace{1cm}v\equiv \frac{k\sqrt{\rho_{m}}}{\sqrt{3}H}\,, \label{eq:avar10}\\
&\lambda\equiv -\dfrac{V_{,\phi}}{k V}\,, \hspace{1cm} \Gamma\equiv \dfrac{V V,_{\phi \phi}}{V_{,\phi}^{2}}\,,\label{eq:avar20}
\end{align}

The constraint equation in terms of these variables is rewritten as
\begin{equation}
    1=\epsilon x^{2}+y^{2}+v^{2},\label{eq:const0}
\end{equation}
and the dark energy equation of state takes the following form $w_{\phi}=-\frac{y^{2}}{\epsilon x^{2}+y^{2}}$.

The presence of variable $\Gamma$ is fundamental if $\lambda$ is not a constant\footnote{The case of constant $\lambda$ is limited to either a constant or an exponential potential.}, so in the general case, the autonomous system reads
\begin{equation}
\begin{cases}
x'=-\frac{3}{2}x y^{2}+\epsilon\sqrt{\frac{3}{2}}\lambda y^{2},&\\
y'=-\sqrt{\frac{3}{2}}\lambda x y+\frac{3}{2}y\left(1-y^{2}\right),&\\
\lambda'=-\sqrt{6}f(\lambda)x,
\label{eq:newdynsys0}
\end{cases}
\end{equation}
where, prime represents derivatives with respect to the number of e-folding $N\equiv\ln a$ and $f(\lambda)=\lambda^{2}\left(\Gamma-1\right)$. The potentials that we want to analyze are resumed in Tab. \ref{tab:potchar}. Here, we can see the explicit expression for $\lambda$ and $\Gamma-1$, i.e., to solve the system, we need to rewrite the function $\Gamma-1$ in terms of the dimensionless variable $\lambda$. We can also determine $\Gamma-1$ as a function of $\lambda$ for the Sugra and AS potentials. In this context, we derive the expressions that relate $u$ and $\lambda$, namely $u_{1,2} =\frac{\pm\sqrt{8 \gamma  \chi +\lambda^2}-\lambda}{4 \gamma }$ for the Sugra potential and $u_{1,2} =\frac{\pm\sqrt{1-A (\lambda-\mu )^2}+B (\lambda-\mu )-1}{\lambda-\mu }$ for the AS potential.  In the specific case of the AS potential where $A =B=0$, the parameter $\Gamma-1$ simplifies to $-(1 + \mu / \lambda)^2/2$. Finally, after solving the system in Eq.~\eqref{eq:newdynsys0}, we can also obtain the evolution of matter by considering Eq.~\eqref{eq:const0}.

\subsubsection{Constraints on dimensionless variables}
The dimensionless variables introduced in Eqs.~\eqref{eq:avar10} and \eqref{eq:avar20} for the analysis of dynamical systems are not well-defined for all values of the potential. To ensure their proper definition, the conditions $V > 0$ and $V_{,\phi} \neq 0$ should be fulfilled for the variables $y$, $\lambda$, and $\Gamma$. These requirements impose specific constraints on the functional forms of the potentials, as detailed below:

For the \emph{Sugra} potential, is required the condition $\chi \neq 2 \gamma k^2 \phi^2$, which implies $\phi \neq \frac{\sqrt{\chi}}{\sqrt{2} \sqrt{\gamma} k}$. The \emph{BCN} potential is well-defined for all values of the parameters $m$ and $l$. The \emph{AS} potential requires $\mu \neq -\frac{2 (B - k \phi)}{A + B^2 - 2 B k \phi + k^2 \phi^2}$, $ A \neq \frac{-\mu B^2 + 2 B k \mu \phi - 2 B - k^2 \mu \phi^2 + 2 k \phi}{\mu}$, and $B \neq \frac{\pm\sqrt{1 - A \mu^2} + k \mu \phi - 1}{\mu}$, leading to $\phi \neq \frac{\pm\sqrt{k^2 - A k^2 \mu^2} + B k \mu + k}{k^2 \mu}$.

The \emph{ULM} potential imposes no restrictions on the parameters $\zeta$ and $n$, and the \emph{IE} potential remains unrestricted as well. For the \emph{Fifth power} potential, $\phi > 0$ holds, while the \emph{Quadratic} and \emph{IHC} potentials require $\phi \neq 0$. Finally, the \emph{CS}, \emph{ISP}, and \emph{Exponential} potentials do not have any additional constraints.

\subsection{Non-minimally coupled cosmology}

In the non-minimally coupled scenario, the Klein-Gordon equation has a source term, see Eq.~\eqref{eq:modKG}, implying the presence of a new dimensionless variable, i.e., $u=k\phi$. Consequently, to find the dynamical behavior of a specific dark energy model, it may be convenient to introduce the following dimensionless variables

\begin{align}
&x\equiv \dfrac{k\dot{\phi}}{\sqrt{6}H}\,, \hspace{1cm} y\equiv \dfrac{k\sqrt{V}}{\sqrt{3}H}\,,\label{eq:avar1}\\
& v\equiv \dfrac{k\sqrt{\rho_{m}}}{\sqrt{3}H}\,, \hspace{1cm} u\equiv k\phi\,.\label{eq:avar2}
\end{align}

Even though these variables are the most prominent in the non-minimally coupled scenario \cite{DAgostino:2018ngy, Carloni:2023egi}, we here seek a more general approach characterizing all the critical points and stability, without fixing a potential \emph{a priori}. To achieve it, we introduce a further term, $\lambda$, in the set of previous variables, including its evolution in the autonomous system. Accordingly, the model will influence the variables themselves plus the Klein-Gordon dynamics as, therein, it furnishes the term $V,_{\phi}$.

Adopting the above strategies, we can now explore how the underlying autonomous systems vary accordingly with a different choice of $U$.

\subsubsection{Einstein theory}

In the context of general relativity, employing the FRW curvature scalar, $R=6(\dot{H}+2H^{2})$, the constraint equation, Eq.~\eqref{eq:FRW11}, turns out to be
\begin{equation}
    1=\epsilon x^{2}+y^{2}+v^{2}+2\sqrt{6}\xi u x+\xi u^{2},\label{eq:constreq}
\end{equation}
and the parameter $s\equiv-\frac{\dot{H}}{H^{2}}$ is given by\footnote{This term resembles the deceleration parameter, widely used in late time cosmology \cite{Mamon:2016dlv} and in cosmographic reconstructions \cite{Visser:2004bf,Capozziello:2021xjw,Luongo:2020aqw,Luongo:2024fww,Bamba:2012cp}, as well as a slow roll parameter, adopted in inflationary scenarios \cite{Linde:2007fr}. Accordingly, it clearly implies how to quantify the corresponding dynamics associated with the autonomous system under exam \cite{Szydlowski:2013sma}.}
\begin{widetext}
\begin{equation}
    s=\frac{1}{1+\xi\left(6\xi-1\right)u^{2}}\left[\frac{3}{\sigma}\left(\epsilon-2\sigma\xi\right)x^{2}+4\sqrt{6}\xi x u+12\xi^{2}u^{2}-3\epsilon \xi \lambda u y^{2}+\frac{3\left(1+w_{m}\right)}{2}v^{2}\right],
\end{equation}
\end{widetext}
where, again, $\lambda=-\frac{V_{,\phi}}{k V}$. Thus, the system becomes
\begin{equation}
\begin{cases}
x'=(s-\frac{3}{\sigma})x+\sqrt{6}\epsilon\xi u\left(s-2\right)+\sqrt{\frac{3}{2}}\epsilon \lambda y^{2},&\\
y'=sy-\sqrt{\frac{3}{2}}\lambda x y,&\\
u'=\sqrt{6}x,&\\
\lambda'=-\sqrt{6}f(\lambda)x.&\label{eq:newdynsys}
\end{cases}
\end{equation}

Again, for what concerns the evolution of matter, it can be derived by solving
Eq.~\eqref{eq:constreq} and, in fact, from it, the dimensionless densities are determined as
\begin{equation}
    \Omega_{m}=v^{2},\hspace{2mm}\Omega_{\phi}=\epsilon x^{2}+y^{2}+2\sqrt{6}\xi u x+\xi u^{2},
\end{equation}
and the effective dark energy equation of state reads
\begin{widetext}
\begin{equation}
    w^{\rm eff}_{\phi}=\frac{\left(3\eta\epsilon x^{2}-3y^{2}-2\xi u\left(-3\sqrt{6}x+6\xi s u-12\xi u+3 \epsilon\lambda y^{2}\right)
    -12\xi x^{2}-4\sqrt{6}\xi x u -\xi\left(-2 s +3\right)u^{2}\right)}{{3\epsilon x^{2}+3y^{2}+6\sqrt{6}\xi x u +3 \xi u^{2}}},
\end{equation}
\end{widetext}
where $\eta=1,0$ for standard and alternative scalar field, respectively.

\subsubsection{Teleparallel and symmetric-teleparallel theories}

As already stressed, we remark that the non-minimally coupled  symmetric-teleparallel picture is investigated imposing the coincident gauge, see Ref. \cite{BeltranJimenez:2017tkd}. Then, these two descriptions of gravity can be unified under the same analysis since the scalar yields the same value in flat FRW, i.e., $T=Q=-6H^{2}$. Here, considering  Eqs.~\eqref{eq:avar1}-\eqref{eq:avar2}, the constraint equation becomes
\begin{equation}
    1=\epsilon x^{2}+y^{2}+v^{2}+\xi u^{2},\label{eq:contrtele}
\end{equation}
and the autonomous system is now rewritten in terms of these variables as
\begin{equation}
\begin{cases}
x'=(s-\frac{3}{\sigma})x+\epsilon \sqrt{6}\xi u+\sqrt{\frac{3}{2}}\epsilon \lambda y^{2},&\\
y'=sy-\sqrt{\frac{3}{2}}\lambda x y,&\\
u'=\sqrt{6}x,&\\
\lambda'=-\sqrt{6}f(\lambda)x,&\label{eq:newdynsys1}
\end{cases}
\end{equation}
where the parameter $s\equiv-\frac{\dot{H}}{H^{2}}$ acquires the form
\begin{equation}
    s=\frac{\frac{3}{\sigma}\epsilon x^{2}-2\sqrt{6}\xi x u +\frac{3}{2}v^{2}}{1-\xi u^{2}}.
\end{equation}
The effective dark energy equation of state becomes
\begin{equation}
    w_{\phi}^{\rm eff}=\frac{\eta \epsilon x^{2}-y^{2}-\xi u^{2}+\frac{2}{3}\xi s u^{2}-4\sqrt{\frac{2}{3}}\xi x u}{\epsilon x^{2}+y^{2}+\xi u^{2}}.
\end{equation}
In addition, in view of the fact that Eq.~\eqref{eq:contrtele} is different from Eq.~\eqref{eq:constreq}, the dimensionless densities are now
\begin{equation}
    \Omega_{m}=v^{2},\hspace{2mm}\Omega_{\phi}=\epsilon x^{2}+y^{2}+\xi u^{2}.
\end{equation}


\section{Stability analysis and physical results}\label{Sec4}

In this section, we conclude with the stability analysis of the dark energy potentials presented above for the different gravity scenarios. Here, we examine the stability of the alternative scalar field description for minimal and non-minimally coupled frameworks, highlighting the differences with the standard scalar field. In this way, we can observe whether the properties of the alternative dark energy alter the cosmological background.

Particularly, we look for the presence of attractors, which are specific critical points characterized by \cite{DAgostino:2018ngy,Carloni:2023egi}:
\begin{itemize}
    \item[-] \textbf{Stable node}, occurring if the eigenvalues of the Jacobian matrix are all negative.
    \item[-] \textbf{Stable spiral}, arising if the real parts of the eigenvalues are negative and the determinant of the matrix computed at the critical point is negative.
\end{itemize}
In the opposite case, the critical point could be a saddle point, i.e., an unstable node or an unstable spiral.

More precisely, following Ref.  \cite{Copeland:2006wr}, we classify

\begin{itemize}
    \item[-] a saddle point to have at least one negative and one positive eigenvalue;
    \item[-] a unstable node to have all positive eigenvalues;
    \item[-] a unstable spiral to have  critical points with eigenvalues that have positive real parts.
\end{itemize}

In all these cases, the point is called hyperbolic, and linear stability analysis is sufficient to determine its stability.

However, if one eigenvalue of the Jacobian matrix has a zero real part, the point is classified as \emph{non-hyperbolic}, and different methods, beyond linear theory, are thus mandatory in order to determine the stability properties, since the linear theory alone fails to be predictive.

To address such conceptual issues, we apply the \emph{center manifold theory} \cite{Boehmer:2011tp}, and in case this approach would fail for specific potentials, we then resort to numerical simulations of the systems in Eqs. \eqref{eq:newdynsys0}, \eqref{eq:newdynsys} and \eqref{eq:newdynsys1}, essentially following the procedure underlined in Ref. \cite{Zhou:2007xp}.

\subsection{Minimally coupled scenario}

In this subsection, we apply the standard stability analysis technique, typically used in the context of a scalar field, to our alternative description, i.e., to quasiquintessence and quasiphantom fields. First, we identify the critical points of the system, and then we compute the eigenvalues at these points to assess stability.

The critical points for the alternative scalar field, minimally coupled with gravity, are detected by setting to zero the derivative with respect to $N$ of the $x$ and $y$ variables, if $\lambda=-\frac{V_{,\phi}}{k V}$ is constant.

Here, to study the dark energy dynamics, we analyze the small perturbations over these variables, i.e.,
\begin{equation}
\left(
\begin{array}{c}
\delta x' \\
\delta y' \\
\end{array}
\right) = {\mathcal J} \left(
\begin{array}{c}
\delta x \\
\delta y \\
\end{array}
\right) \,,
\label{eq:pert0}
\end{equation}
where $\mathcal{J}$ is the Jacobian matrix.

Specifically, the small perturbations $\delta x$ and $\delta y$ are determined by the Jacobian matrix
\begin{equation}
\mathcal{J}=\left( \begin{array}{cc}
\frac{\partial x'}{\partial x}& \frac{\partial x'}{\partial y}\\
\frac{\partial y'}{\partial x}& \frac{\partial y'}{\partial y}\\
\end{array} \right)_{(x=x_c,y=y_c)}\,,
\label{eq:J0}
\end{equation}
in which
$\textbf{X}_{c}=\left(x_{c},y_{c}\right)$ represents the critical point, indicated by the subscript $c$.

However, if $\lambda$ is no longer a constant, it is necessary to include it as well. Hence, Eq.~\eqref{eq:pert0} becomes
\begin{equation}
\left(
\begin{array}{c}
\delta x' \\
\delta y' \\
\delta \lambda' \\
\end{array}
\right) = {\mathcal J} \left(
\begin{array}{c}
\delta x \\
\delta y \\
\delta \lambda\\
\end{array}
\right) \,,
\label{eq:pertnew0}
\end{equation}
with $\mathcal{J}$ given by
\begin{equation}
\mathcal{J}=\left( \begin{array}{ccc}
\frac{\partial x'}{\partial x}& \frac{\partial x'}{\partial y} & \frac{\partial x'}{\partial \lambda} \\
\frac{\partial y'}{\partial x}& \frac{\partial y'}{\partial y} & \frac{\partial y'}{\partial \lambda}\\
\frac{\partial \lambda'}{\partial x}& \frac{\partial \lambda'}{\partial y} & \frac{\partial \lambda'}{\partial \lambda}\\
\end{array} \right)_{(x=x_c,y=y_c,\lambda=\lambda_{c})}\,,
\label{eq:newJ0}
\end{equation}
where, this time, the critical point turns out to be
$\textbf{X}_{c}=\left(x_{c},y_{c},\lambda_{c}\right)$.

Working this way, we then developed a general approach for the alternative scalar field scenarios, namely a treatment that turns out to be  valid for \emph{all the types of involved potentials}.

In addition, the above technique allows us to highlight the main expected differences among the standard and alternative descriptions of the scalar field \cite{Bahamonde:2017ize}.

Precisely, the main difference is that, in the alternative scalar field case, there are two critical points where $\omega_{\phi}=0$, see Tab. \ref{tab:criticalmin}. This implies that, as dynamics ends, dark energy mimes dust, physically interpreting the quasiquintessence field to behave as a \emph{unified dark energy-dark matter fluid} at least for what concerns a pure dynamical perspective, as also confirmed in Ref. \cite{Gao:2009me}, adopting a different path. Afterwards, to investigate the stability of the systems, it behooves us to study the eigenvalues computed at the critical points, as reported in Tab. \ref{tab:stabilitymin}.

\begin{table*}
\centering
\resizebox{0.9\textwidth}{!}{
\footnotesize
\setlength{\tabcolsep}{0.2em}
\renewcommand{\arraystretch}{2}
\begin{tabular}{|c|c|c|c|p{2.5cm}|p{2.5cm}|p{2.5cm}|p{2.5cm}|}
\hline\hline
Point & x & y & $\lambda$ & Existence & $\omega_{\phi}$ & Accel. & $\Omega_{\phi}$\\
\hline\hline
$P_{0}$ & 0 & 0 & Any & Always & $0$ & No & 0\\
$P_{1}$ & Any & 0 & $\lambda_{*}$ & $\lambda_{*}=0$ & $0$ & No & $\epsilon x^2$\\
$P_{2}$ & Any & 0 & $\lambda_{*}$ & $\lambda_{*}\neq 0$ & $0$ & No & $\epsilon x^2$\\
$P_{3}$ & $\sqrt{\frac{2}{3}}\lambda_{*}$ & $\sqrt{1-\frac{2}{3}\lambda_{*}^{2}}$ & $\lambda_{*}$ & $\lambda_{*}^{2}<\frac{3}{2}$ & $-1+\frac{4}{3}\lambda^{2}_{*}$ & $\lambda^{2}_{*}<\frac{1}{2}$ & $1$\\
$P_{4}$ & $0$ & $1$ & $0$ & Always & $-1$ & Yes & $1$\\
\hline\hline
\end{tabular}}
\caption{The critical points of the system representing the alternative scalar field, minimally coupled to gravity, are given along with their existence conditions and cosmological values. The value $\lambda_{*}$ is any zero of the function $f(\lambda)=\lambda^{2}(\Gamma-1)$.}
\label{tab:criticalmin}
\end{table*}

 \begin{table*}
    \centering
      \resizebox{0.9\textwidth}{!}{
\footnotesize
\setlength{\tabcolsep}{0.2em}
\renewcommand{\arraystretch}{2}
\begin{tabular}{|c|c|c|c|c|}
\hline\hline
Point  & Eigenvalues & Hyperbolicity & Stability \\
\hline\hline
\multirow{3}*{$P_0$} & \multirow{3}*{$\{0,0,\frac{3}{2}\}$} & \multirow{3}*{No} & \multirow{2}*{Unstable} \\ & & & \\
\multirow{3}*{$P_1$} & \multirow{3}*{$\{0,0,\frac{3}{2}\}$} & \multirow{3}*{No} & \multirow{2}*{Unstable} \\ & & & \\
\multirow{4}*{$P_2$} & \multirow{4}*{$\{0,-\sqrt{6}\lambda^{2}_{*}\Gamma'_{*}x,\frac{3}{2}-\sqrt{\frac{3}{2}}\lambda_{*}x\}$} & \multirow{4}*{No}& \multirow{2}*{Indeterminate if $\frac{3}{2}-\sqrt{\frac{3}{2}}\lambda_{*}x <0$ and $\lambda^{2}_{*}\Gamma'_{*}x\geq0$}
\\ & & & \multirow{2}*{Unstable if $\frac{3}{2}-\sqrt{\frac{3}{2}}\lambda_{*}x>0$ and $\lambda^{2}_{*}\Gamma'_{*}x<0$} \\
 & & & \multirow{2}*{Saddle if $\frac{3}{2}-\sqrt{\frac{3}{2}}\lambda_{*}x<0$ and $-\lambda^{2}_{*}\Gamma'_{*}x>0$ or viceversa} \\ & & & \\
\multirow{5}*{$P_3$} & \multirow{5}*{$\{-2\lambda^{3}_{*}\Gamma'_{*},-\frac{3}{2}+\lambda^{2}_{*},-3+2\lambda^{2}_{*}\}$} & \multirow{5}*{Yes} & \multirow{3}*{Stable if $-\sqrt{\frac{3}{2}}<\lambda_{*}<0$, $\Gamma_{*}'<0$ or $0<\lambda_{*}<\sqrt{\frac{3}{2}}$, $\Gamma_{*}'>0$} \\ & & & \multirow{3}*{Unstable if $\lambda_{*}<-\sqrt{\frac{3}{2}}$, $\Gamma_{*}'>0$ or $\lambda_{*}>\sqrt{\frac{3}{2}}$, $\Gamma_{*}'<0$} \\ & & & \multirow{3}*{Saddle otherwise}\\

& & &\\

\multirow{4}*{$P_4$} & \multirow{4}*{$\left\{-3,-\frac{3}{4}-\sqrt{\frac{9}{16}-3 f(0)},-\frac{3}{4}+\sqrt{\frac{9}{16}-3 f(0)}\right\}$} & \multirow{4}*{Yes if $f(0)\neq0$}& \multirow{3}*{Stable if $f(0)>0$} \\ & & & \multirow{3}*{Saddle if $f(0)<0$} \\ & & & \\
\hline\hline
\end{tabular}}
\caption{Stability properties of the critical points for the alternative scalar field minimally coupled to gravity.}
\label{tab:stabilitymin}
\end{table*}

We then observe that the non-hyperbolicity prevents us from determining the stability of points $P_2$ and $P_4$ when $\frac{3}{2}-\sqrt{\frac{3}{2}}\lambda_{*}x<0$ and $\lambda^{2}_{*}\Gamma'_{*}x\geq0$, and $f(0)=0$, respectively. Here, $\lambda_{*}$ is any zero of function $f(\lambda)=\lambda^{2}(\Gamma-1)$ and $\Gamma'_{*}$ is the derivative of $\Gamma$ with respect $\lambda$ computed in $\lambda_{*}$.

\begin{figure*}
\centering
\subfigure[AS potential with free parametes set as $A=0$, $B=0$ and $\mu=3$.\label{P2ASMIN}]
{\includegraphics[height=0.4\hsize,clip]{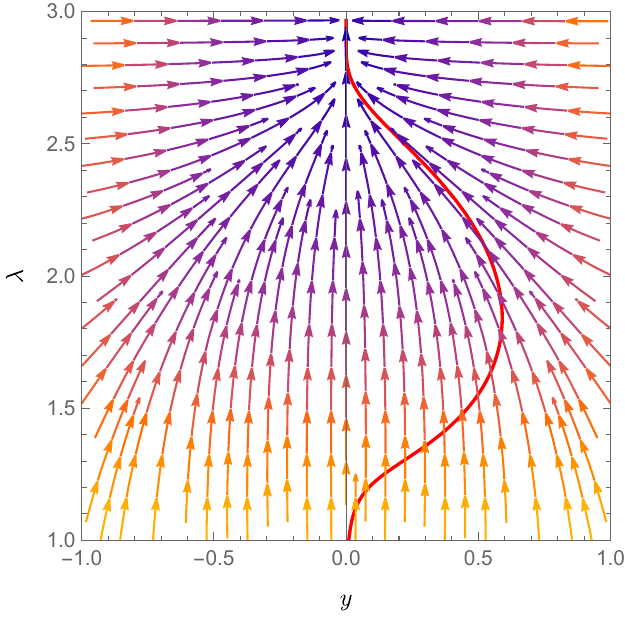}}
\subfigure[BCN potential with free parameters set as $m=\frac{1}{2}$ and $l=6$.\label{P2BCNMIN}]
{\includegraphics[height=0.4\hsize,clip]{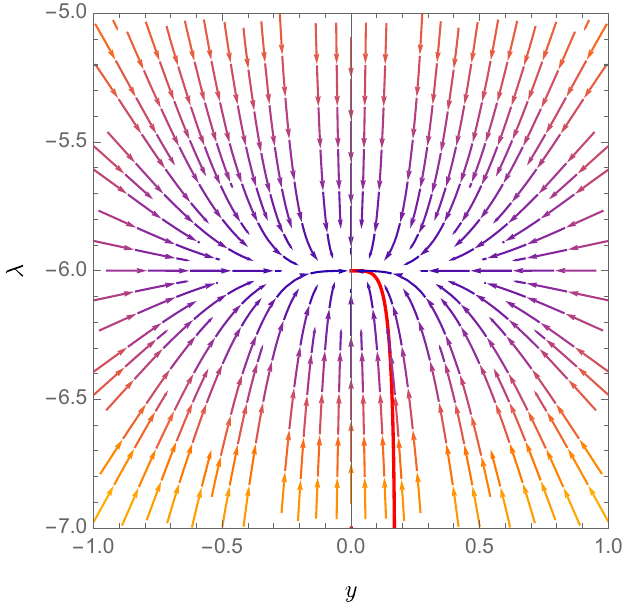}}
\subfigure[ULM potential with free parameters set as $n=-1$ and $\zeta=2$.\label{P2ULMMIN}]
{\includegraphics[height=0.4\hsize,clip]{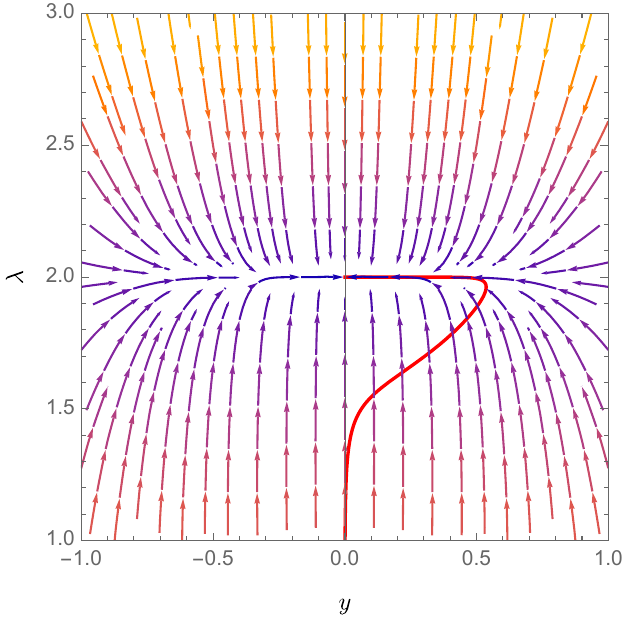}}
\caption{Phase-space trajectories on $y-\lambda$ plane for AS, BCN and ULM potentials, where the red lines represent the solution of the dynamical systems. The initial conditions are set as $x_{\rm in}=0.03$, $y_{\rm in}=10^{-3}$ for all potentials, with $\lambda_{\rm in}=-7$ for the BCN potential and $\lambda_{\rm in}=1$ for the AS and ULM potentials. From the numerical analysis, we observe that point $P_{2}$ is stable, i.e., an attractor, in all cases.}
\end{figure*}

\begin{figure*}
\centering
{\includegraphics[height=0.4\hsize,clip]{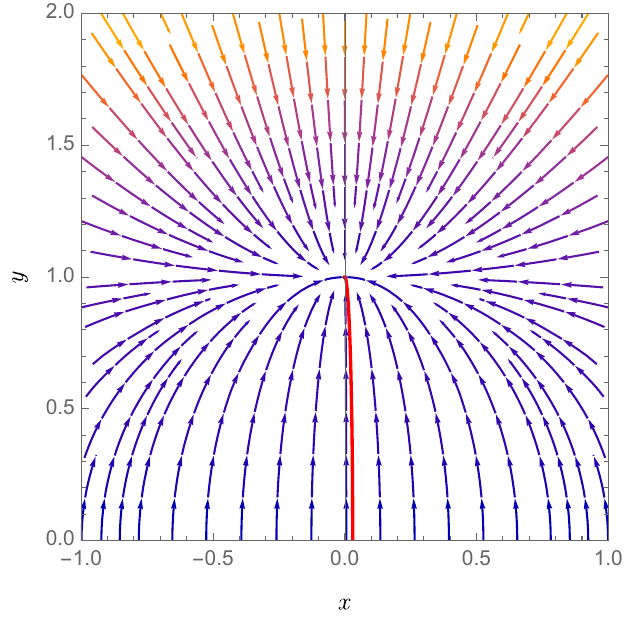}}
\caption{Phase-space trajectories on $x-y$ plane for IE and CS potentials, where the red line represents the solutions of the dynamical systems. The free parameter for the CS potential is  selected as $\tau=1$. The initial conditions are set as $x_{\rm in}=0.03$, $y_{\rm in}=10^{-3}$, and $\lambda_{\rm in}=10^{-10}$. Under these conditions, the solutions for the IE and CS potentials are identical, so they are represented on a single graph. Point $P_{4}$ is stable, i.e., an attractor, in both cases.}
\label{P4Inv}
\end{figure*}

Thus, to determine whether these points can be attractors for the system, we need to select a specific potential and assess the stability of the model under examination by applying center manifold theory or numerical analysis.

For what concerns the critical point $P_2$, the center manifold theory does not provide useful information for analyzing the behavior around the critical point, and we have to rely on numerical methods to establish whether the point is an attractor. It is important to note that if one or both of the eigenvalues are negative, the point cannot be an attractor and is classified as an unstable or saddle point, as for the CS model. The condition on the free parameter for this potential ensures that at least one eigenvalue is positive. In contrast, for the Sugra potential, $\lambda_{*}$ is imaginary, which means that this critical point does not exist.

In addition, since the dimensionless scalar field density is modeled by $\Omega_{\phi}=\epsilon x^{2}$, the phantom field appears unphysical as it provides a negative density value. For the sake of completeness, negative densities in cosmology are not fully-excluded yet. Examples are furnished by Refs. \cite{Farnes:2017gbf,Visinelli:2019qqu, Menci:2024rbq}. However, the conservative approach is to work positive densities only, as claimed throughout this work.

Thus, the numerical techniques is applied for BCN, AS and ULM potentials imposing $v_{\rm in}^{2}= 0.999$ at $z = 10^{2}$ as a realistic initial condition for matter, see Refs. \cite{Carloni:2023egi, Xu:2012jf, DAgostino:2018ngy}.

The dynamical system is then solved for each potential by setting $x_{\rm in}$, $y_{\rm in}$, and $\lambda_{\rm in}$ to ensure convergence to the critical point itself, and the results are displayed in Figs. \ref{P2ASMIN}, \ref{P2BCNMIN} and \ref{P2ULMMIN}.

Thus, we can conclude that, for these potentials, the critical point, $P_{2}$, behaves as an attractor. This indicates a scenario where, within the critical point, dark energy may exhibit dark matter characteristics, i.e., dust-like properties, since the corresponding equation of state tends to vanish there.

At this stage, we subsequently analyze the critical point $P_{4}$. In the first instance, we observe that, if $\text{det}\mathcal{J}<0$ and $\frac{9}{16}-3f(0)<0$, the point is a stable spiral, yielding an attractor behavior. Convsersely, if $f(0)=0$, the critical point is non-hyperbolic and, so, we require alternative techniques to study the stability. We first apply the center manifold theory, and the critical point point is an attractor if $\Gamma(0)>1$, as reported in \ref{CMT}. However, if $\Gamma(0)$ diverges, we can not deduce anything about this point using the center manifold theory and we recur to numerical computation. This is the case of IE and CS potentials. Again, we use $v_{\rm in}^{2}=0.999$ at $z=10^{2}$ as realistic initial conditions to obtain the numerical solution, and the results are represented in Fig. \ref{P4Inv}.
Then, we can conclude that, only for BCN, ULM and IHC, the critical point $P_{4}$ is a saddle point, due to the fact that $f(0)<0$. For the other potentials, the critical point is an attractor in which dark energy behaves as a cosmological constant, dominating the universe. In all the scenarios, conventionally to let the field evolve up to future times, we single out $N_f=\ln(a)\simeq 10;25;50$, where $N_f=0$ indicates today, i.e., $a=1$.

\subsection{Non-minimally coupled scenario}

In this subsection, we analyze the stability of systems involving non-minimally coupled standard and alternative scalar field models, proposing a generalized approach to derive critical points without indicating the potential a priori. The use of gauge in the geometrical trinity of gravity is specifically associated with the case of $\xi = 0$. We remark that the non-minimal coupling, i.e., $\xi \neq 0$, in principle, suggests that the gauge is no longer valid \emph{a priori} and may differ. However, in alignment with Refs.\cite{Xu:2012jf,Hrycyna:2008gk,Carloni:2023egi,DAgostino:2018ngy,Lu:2019hra,Ghosh:2023amt}, we use the same gauge, intentionally departing from the original trinity of gravity. As a result, we refer to this as the ``modified trinity of gravity'', where the latter is due to the non-minimal coupling.

 \begin{table*}
    \centering
      \resizebox{1.0\textwidth}{!}{
\footnotesize
\setlength{\tabcolsep}{0.2em}
\renewcommand{\arraystretch}{2}
\begin{tabular}{|c|c|c|c|c|p{4.0cm}|p{2.0cm}|p{1.5cm}|p{2.5cm}|}
\hline
\hline
Point & x & y & u & $\lambda$ & Existence & $\omega^{\rm eff}_{\phi}$ & Accel. & $\Omega^{\rm eff}_{\phi}$\\
\hline\hline
   \multirow{2}{*}{$P_{0}$} &\multirow{2}{*}{$0$} & \multirow{2}{*}{$0$} & \multirow{2}{*}{$\pm\sqrt{\frac{1}{\xi}}$} & \multirow{2}{*}{Any} &\multirow{2}{*}{$\xi>0$}  & \multirow{2}{*}{$\frac{1}{3}$} &\multirow{2}{*}{No} &\multirow{2}{*}{$1$}\\
\multirow{3}{*}{$P_{1}$} &\multirow{3}{*}{$0$} &\multirow{3}{*}{$0$} &\multirow{3}{*}{$0$} &\multirow{3}{*}{Any} &\multirow{3}{*}{Always} &\multirow{3}{*}{$\frac{0}{0}$} &\multirow{3}{*}{Indeter.} &\multirow{3}{*}{$0$}\\
\multirow{4}{*}{$P_{2}$} &\multirow{4}{*}{$0$} &\multirow{4}{*}{$0< y^{2} <1$} &\multirow{4}{*}{$\pm\sqrt{\frac{1-y^2}{\xi +8 \xi ^2 (\epsilon -1)}}$} &\multirow{4}{*}{$\pm\frac{4 \xi}{y^{2}}  \sqrt{\frac{1-y^2}{\xi +8 \xi ^2 (\epsilon -1)}}$} &\multirow{4}{*}{for $\xi>0$ if $\epsilon=1$} & \multirow{4}{*}{$\frac{-8 \xi  (\epsilon -1)-1}{8 y^2 \xi  (\epsilon -1)+1}$} &\multirow{4}{*}{-} &\multirow{4}{*}{$\frac{8 y^2 \xi  (\epsilon -1)+1}{8 \xi  (\epsilon -1)+1}$}\\
& & & & &\multirow{3}{*}{for $0<\xi <\frac{1}{16}$ if $\epsilon=-1$}& & \multirow{3}{*} &  \\
\multirow{4}{*}{$P_{3}$} &\multirow{4}{*}{$0$} &\multirow{4}{*}{$y^{2}>1$} &\multirow{4}{*}{$\pm\sqrt{\frac{1-y^2}{\xi +8 \xi ^2 (\epsilon -1)}}$} &\multirow{4}{*}{$\pm\frac{4 \xi}{y^{2}}  \sqrt{\frac{1-y^2}{\xi +8 \xi ^2 (\epsilon -1)}}$} &\multirow{4}{*}{for $\xi<0$ if $\epsilon=1$} & \multirow{4}{*}{$\frac{-8 \xi  (\epsilon -1)-1}{8 y^2 \xi  (\epsilon -1)+1}$} &\multirow{4}{*}{-} &\multirow{4}{*}{$\frac{8 y^2 \xi  (\epsilon -1)+1}{8 \xi  (\epsilon -1)+1}$}\\
& & & & &\multirow{3}{*}{for $\xi<0$ or $\xi>\frac{1}{16}$ if $\epsilon=-1$}& & &  \\
\multirow{3}{*}{$P_{4}$} &\multirow{3}{*}{$0$} &\multirow{3}{*}{$1$} &\multirow{3}{*}{$0$} &\multirow{3}{*}{$0$} &\multirow{3}{*}{Always} &\multirow{3}{*}{$-1$} &\multirow{3}{*}{Yes} &\multirow{3}{*}{$1$}\\
& & & & & & & & \\
\hline\hline
\end{tabular}}
\caption{Critical points of the system with existence condition and cosmological values for non-minimal coupling cosmology. These points are the same for the standard and alternative scalar field descriptions. For the points $P_2$ and $P_3$, the acceleration condition is determined by setting the correct value of $\xi$ if $\epsilon=-1$.}\label{tab:criticalnonmin}
\end{table*}

 \begin{table*}
    \centering
      \resizebox{0.9\textwidth}{!}{
\footnotesize
\setlength{\tabcolsep}{0.2em}
\renewcommand{\arraystretch}{2}
\begin{tabular}{|c|c|c|c|}
\hline\hline
Point  & Eigenvalues & Hyperbolicity & Stability \\
\hline\hline
\multicolumn{4}{|c|}{Standard scalar field}\\
\hline\hline
\multirow{3}*{$P_0$} & \multirow{3}*{$\{0,\frac{3}{2},\frac{1}{4}(-3-\sqrt{9-48\epsilon \xi}),\frac{1}{4}(-3+\sqrt{9-48\epsilon \xi})\}$} & \multirow{3}*{No} & \multirow{3}*{Saddle} \\ & & & \\
\multirow{3}*{$P_1$} & \multirow{3}*{$\{-1,0,2,\epsilon\}$} & \multirow{3}*{No} & \multirow{3}*{Saddle} \\ & & & \\
\multirow{4}*{$P_4$} & \multirow{4}*{$\{-3,0,\frac{1}{2}(-3-\sqrt{9-48\epsilon \xi}),\frac{1}{2}(-3+\sqrt{9-48\epsilon \xi})\}$} & \multirow{4}*{No}& \multirow{3}*{Saddle if $\sqrt{9-48\epsilon \xi}>3$}\\ & & & \multirow{3}*{Indeterminate otherwise}\\ & & & \\
\hline\hline
\multicolumn{4}{|c|}{Alternative scalar field}\\
\hline\hline
\multirow{3}*{$P_0$} & \multirow{3}*{$\{0,\frac{3}{2},-i\sqrt{3\epsilon\xi},i\sqrt{3\epsilon\xi})\}$} & \multirow{3}*{No} & \multirow{3}*{Unstable} \\ & & & \\
\multirow{3}*{$P_1$} & \multirow{3}*{$\{0,2,\frac{1}{4} \left(-\sqrt{4 \epsilon ^2+20 \epsilon +1}+2 \epsilon +1\right),\frac{1}{4} \left(\sqrt{4 \epsilon ^2+20 \epsilon +1}+2 \epsilon +1\right)\}$} & \multirow{3}*{No} & \multirow{3}*{Saddle} \\ & & & \\
\multirow{4}*{$P_4$} & \multirow{4}*{$\{-3,0,\frac{1}{2}(-3-\sqrt{9-192\epsilon \xi}),\frac{1}{2}(-3+\sqrt{9-192\epsilon \xi})\}$} & \multirow{4}*{No}& \multirow{3}*{Saddle if $\sqrt{9-192\epsilon \xi}>3$}\\ & & & \multirow{3}*{Indeterminate otherwise}\\ & & & \\
\hline\hline
\end{tabular}}
\caption{Stability properties for the critical points of the non-minimal coupled standard scalar field system. The eigenvalues of $P_{2}$ and $P_{3}$ are of the same kind as $P_{4}$, but they depend on the explicit form of $y$ and $\Gamma$.}
\label{tab:stabilitynonmin}
\end{table*}

 \begin{table*}
    \centering
      \resizebox{0.9\textwidth}{!}{
\footnotesize
\setlength{\tabcolsep}{0.4em}
\renewcommand{\arraystretch}{2}
\begin{tabular}{|c|c|c|c|c|p{2.5cm}|p{2.5cm}|p{2.5cm}|p{2.5cm}|}
\hline
\hline
Point & x & y & u & $\lambda$ & Existence & $\omega^{\rm eff}_{\phi}$ & Accel. & $\Omega^{\rm eff}_{\phi}$\\
\hline\hline
\cline{1-9}
\hline\hline
   $P_{0}$ & 0 & 0 & 0 & Any & Always & $\frac{0}{0}$ & Indeterminate & 0\\
$P_{1}$ & 0 & $0<y^{2}< 1$ &  $\pm\sqrt{\frac{1-y^2}{\xi }}$ & $\mp\frac{2 \xi}{y^{2}}\sqrt{\frac{1-y^2}{\xi}}$ & $\xi >0$ & $-1$ & Yes & $1$\\
$P_{2}$ & 0 & $y^{2}>1$ &  $\pm\sqrt{\frac{1-y^2}{\xi }}$ & $\mp\frac{2 \xi}{y^{2}}\sqrt{\frac{1-y^2}{\xi}}$ & $\xi <0$ & $-1$ & Yes & $1$\\
$P_{3}$ & $0$ & $1$ & $0$ & $0$ & Always & $-1$ & Yes & $1$\\

\hline\hline

\end{tabular}}
\caption{Critical points of the teleparallel and symmetric-teleparallel dark energy systems with existence condition and cosmological values. These points are the same for the standard and alternative scalar field descriptions.}\label{tab:criticaltesymte}
\end{table*}

\begin{table*}
\centering
\resizebox{0.9\textwidth}{!}{
\footnotesize
\setlength{\tabcolsep}{0.2em}
\renewcommand{\arraystretch}{1.5}
\begin{tabular}{|c|p{7cm}|p{4cm}|p{4cm}|}
\hline\hline
Point  & Eigenvalues & Hyperbolicity & Stability \\
\hline\hline
\multicolumn{4}{|c|}{Standard scalar field}\\
\hline\hline
\multirow{2}*{$P_0$} & \multirow{2}*{$\{0,\frac{3}{2},\frac{1}{4} \left(-3-\sqrt{9+96 \xi  \epsilon}\right),\frac{1}{4} \left(-3+\sqrt{9+96 \xi  \epsilon}\right)\}$} & \multirow{2}*{No} & Saddle \\
& & & \\

\multirow{3}*{$P_3$} & \multirow{3}*{$\{-3,0,\frac{1}{2} \left(-3- \sqrt{9+24\xi  \epsilon}\right),\frac{1}{2} \left(-3+\sqrt{9+24\xi  \epsilon}\right)\}$} & \multirow{3}*{No} & Saddle if $\sqrt{9+24\xi  \epsilon}>3$ \\
& & & Indeterminate otherwise \\
& & & \\
\hline\hline
\multicolumn{4}{|c|}{Alternative scalar field}\\
\hline\hline
\multirow{2}*{$P_0$} & \multirow{2}*{$\{0,\frac{3}{2},\sqrt{6\epsilon\xi},-\sqrt{6\epsilon\xi}\}$} & \multirow{2}*{No} & Saddle \\
& & & \\

\multirow{3}*{$P_3$} & \multirow{3}*{$\{-3,0,\frac{1}{4} \left(-3-\sqrt{3+96 \xi  \epsilon}\right),\frac{1}{4} \left(-3+\sqrt{3+96 \xi  \epsilon}\right)\}$} & \multirow{3}*{No} & Saddle if $\sqrt{3+96 \xi  \epsilon}>3$ \\
& & & Indeterminate otherwise \\
& & & \\
\hline\hline
\end{tabular}}
\caption{Stability properties for the critical points of the teleparallel and symmetric-teleparallel dark energy systems. The eigenvalues of $P_{1}$ and $P_{2}$ are of the same kind as $P_{3}$, but they depend on the explicit form of $y$ and $\Gamma$.}
\label{tab:stabilityTE}
\end{table*}

Here, we obtain the same critical points for both the standard and alternative scalar field descriptions, indicating that the presence of coupling between dark energy and gravity unifies these two different types of scalar field. In particular, the dust-like behavior of the alternative scalar field disappears in favor of a typical dark energy description. To study the stability of these point, we study again the behavior of small perturbations. This time, the system has an additional variable due to the source term in the Klein-Gordon equation, so we consider the following linear evolution of perturbations
\begin{equation}
\left(
\begin{array}{c}
\delta x' \\
\delta y' \\
\delta u' \\
\delta \lambda'\\
\end{array}
\right) = {\mathcal J} \left(
\begin{array}{c}
\delta x \\
\delta y \\
\delta u \\
\delta \lambda
\end{array}
\right) \,,
\label{eq:pert}
\end{equation}
with the Jacobian matrix determined as
\begin{equation}
 \label{matJ2}
\mathcal{J}=\left( \begin{array}{cccc}
\frac{\partial x'}{\partial x}& \frac{\partial x'}{\partial y} & \frac{\partial x'}{\partial u}&
\frac{\partial x'}{\partial \lambda}\\
\frac{\partial y'}{\partial x}& \frac{\partial y'}{\partial y} & \frac{\partial y'}{\partial u}&
\frac{\partial \lambda'}{\partial \lambda}\\
\frac{\partial u'}{\partial x}& \frac{\partial u'}{\partial y} & \frac{\partial u'}{\partial u}&
\frac{\partial \lambda'}{\partial \lambda}\\
\frac{\partial \lambda'}{\partial x}& \frac{\partial \lambda'}{\partial y} & \frac{\partial \lambda'}{\partial u}&
\frac{\partial \lambda'}{\partial \lambda}\\
\end{array} \right)_{(x=x_c,y=y_c,u=u_{c},\lambda=\lambda_{c})}.
\end{equation}

The critical points and corresponding eigenvalues for non-minimally coupled and teleparallel/symmetric-teleparallel dark energy are presented Tabs. \ref{tab:criticalnonmin}-\ref{tab:stabilitynonmin} and \ref{tab:criticaltesymte}-\ref{tab:stabilityTE}, respectively. From these tables, we observe that the main difference between the gravity frameworks lies in the existence of a saddle critical point where dark energy behaves as radiation in non-minimally coupled dark energy, which is absent in both teleparallel and symmetric-teleparallel dark energy. This behavior is similar to the results found in Ref. \cite{Szydlowski:2013sma}, albeit apparently it does not furnish new physics but just a convergence point. {

Furthermore, as anticipated, the tables reveal no distinction between the standard and alternative descriptions of scalar field at critical points, since as stated above, the non-minimal coupling tends to hide the different characteristics of the potentials.

However, a subtle difference emerges when calculating the eigenvalues at the critical points, as shown in Tabs. \ref{tab:stabilitynonmin} and \ref{tab:criticaltesymte}. The linear stability analysis confirms that all the critical points are non-hyperbolic.  The candidate attractor points are $P_2$, $P_3$, and $P_4$ for non-minimally coupled dark energy, and $P_2$ and $P_3$ for teleparallel and symmetric-teleparallel dark energy. Due to the large number of free parameters in this case, we perform a numerical analysis to determine the behavior of the dark energy systems.

In this respect, to numerically solve the systems, it is convenient to not consider the equation related to $\lambda'$. In fact, by explicitly expressing the potential, we can rewrite $\lambda$ in terms of $u$, allowing us to solve each system with fewer variables. To evaluate the dynamics of dark energy, we assume that the universe is initially dominated by matter, setting $v^{2}_{\rm in}=0.999$ at $z=10^{2}$ \cite{Carloni:2023egi, Xu:2012jf, DAgostino:2018ngy}. With this assumption, we let the system evolve by changing the potential and the initial conditions associated with it. In particular, the initial conditions are chosen in the range $y_{\rm in}\in [10^{-4},10^{-2}]$ and $u_{\rm in} \in [10^{-4},10^{-1}]$ to reach the convergence. Then, the value $x_{\rm in}$ is determined by solving the constraint equation. Among all the models, the only one that does not admit negative values of the field (i.e., negative values of $u$) is the fifth potential, as for such values, the potential becomes negative, leading to an imaginary dimensionless variable $y$.

Finally, in Fig.~\ref{StRmodels} are prompted our findings for non-minimal coupled standard scalar field, while in Fig.~\ref{StTQmodels} are displayed the results for teleparallel and symmetric-teleparallel standard scalar field in the coincident gauge. All these results are also valid for the alternative scalar field description, suggesting that the coupling between the scalar field and gravity masks the differences between the two types of scalar fields, as it dominates over the potentials themselves for given values of the field.


\section{Growth of matter perturbations}\label{Sec5}

In this section, we examine the influence of corrections introduced by our dark energy models within the non-minimally coupled theories of gravity on the dynamics of the universe at high redshifts. Specifically, we focus on density perturbations, which provide a robust framework for assessing the efficacy of cosmological models in describing the early universe\cite{Dunsby:2016lkw}.

When dark energy is non-minimally coupled to gravity, the growth of matter perturbations is governed by the equation
\begin{equation}
    \ddot{\delta} + 2H\dot{\delta} - 4\pi G_{\rm{eff}}\rho_{m}\delta = 0,\label{eq:matper1}
\end{equation}
where the density contrast is defined as $\delta \equiv \delta_{m}/\rho_{m}$, and the effective gravitational constant is given by $G_{\rm{eff}} = \alpha G_{N}$. The parameter $\alpha$ is determined by the specific theory of gravity under consideration. Notably, $\alpha = 1$ in the minimally coupled regime, while in $R$ and $T$ gravity, it takes the following forms:

\begin{itemize}
    \item[-] For Einstein theory \cite{Fu:2010zza, Boisseau:2000pr}
    \begin{equation}
        \alpha = \frac{1}{\mathcal{F}_{R}}\left(\frac{2\epsilon \mathcal{F}_{R} + 4\mathcal{F}^{2}_{R\phi}}{2\epsilon \mathcal{F}_{R} + 3\mathcal{F}^{2}_{R\phi}}\right) \simeq \frac{1}{\mathcal{F}_{R}} = \frac{1}{1 - \xi u^{2}},
    \end{equation}
    where the simplification holds as $\mathcal{F}_{R\phi}^{2} = 4k^{2}\xi^{2} u^{2} \ll 1$.

    \item[-] For teleparallel theory\cite{Geng:2012vn, Golovnev:2018wbh}
    \begin{equation}
        \alpha = \frac{1}{1 - \xi u^{2}}\left(1 - \frac{3\epsilon x^{2}}{1 - \xi u^{2}}\right).\label{GeffT}
    \end{equation}
\end{itemize}

Here, Eq.~\eqref{GeffT} is also applicable to describe the effective gravitational constant in the symmetric-teleparallel theory, as we adopt the coincident gauge \cite{BeltranJimenez:2019tme, Khyllep:2021pcu}.

To analyze the perturbation equation, we rewrite Eq.~\eqref{eq:matper1} in terms of the growth index $f \equiv \frac{d \ln \delta}{dN}$, where $N = \ln a$. This reformulation leads to
\begin{equation}
    \frac{df}{dN} + f^{2} + \frac{1}{2}\left(1 - \frac{d\ln E}{dN}\right)f = \frac{3}{2}\frac{G_{\rm eff}}{G_{N}}\Omega_{m},
\end{equation}
with $E = H/H_{0}$ defined as
\begin{equation}
    E = \sqrt{\Omega_{m0}a^{-3} + \Omega_{\phi 0}\exp\left(3 \int_{a}^{1} \frac{1 + \omega^{\rm eff}_{\phi}}{a'} da'\right)},
\end{equation}
where $\Omega_{\phi 0} = 1 - \Omega_{m 0}$, the subscript $0$ refers to our time and $\omega^{\rm eff}_{\phi}=\omega_{\phi}$ in the minimally coupled scenario. We impose the realistic boundary condition $f(a_{\rm LSS}) = 1$, where $a_{\rm LSS} = (1 + z_{\rm LSS})^{-1}$ represents the scale factor at the last scattering surface, approximated as $z_{\rm LSS} \approx 1089$. The growth history of our models is compared to that of the $\Lambda$CDM model, in order to underline the differences. In particular, we define a region of $\pm 10\%$ deviations from $\Lambda$CDM, as shown in Figs.~\ref{PEDERALL}, \ref{PEDETALL}, and \ref{PEDEALL}. We observe that the only potential fitting within this $\pm 10\%$ range is the quasiphantom field minimally coupled to gravity with fifth power and quadratic potentials, while the other models exceed this limit.

\begin{figure*}
\centering
\subfigure[Growth index for the non-minimally coupled quintessence.\label{PEDER}]
{\includegraphics[height=0.52\hsize,clip]{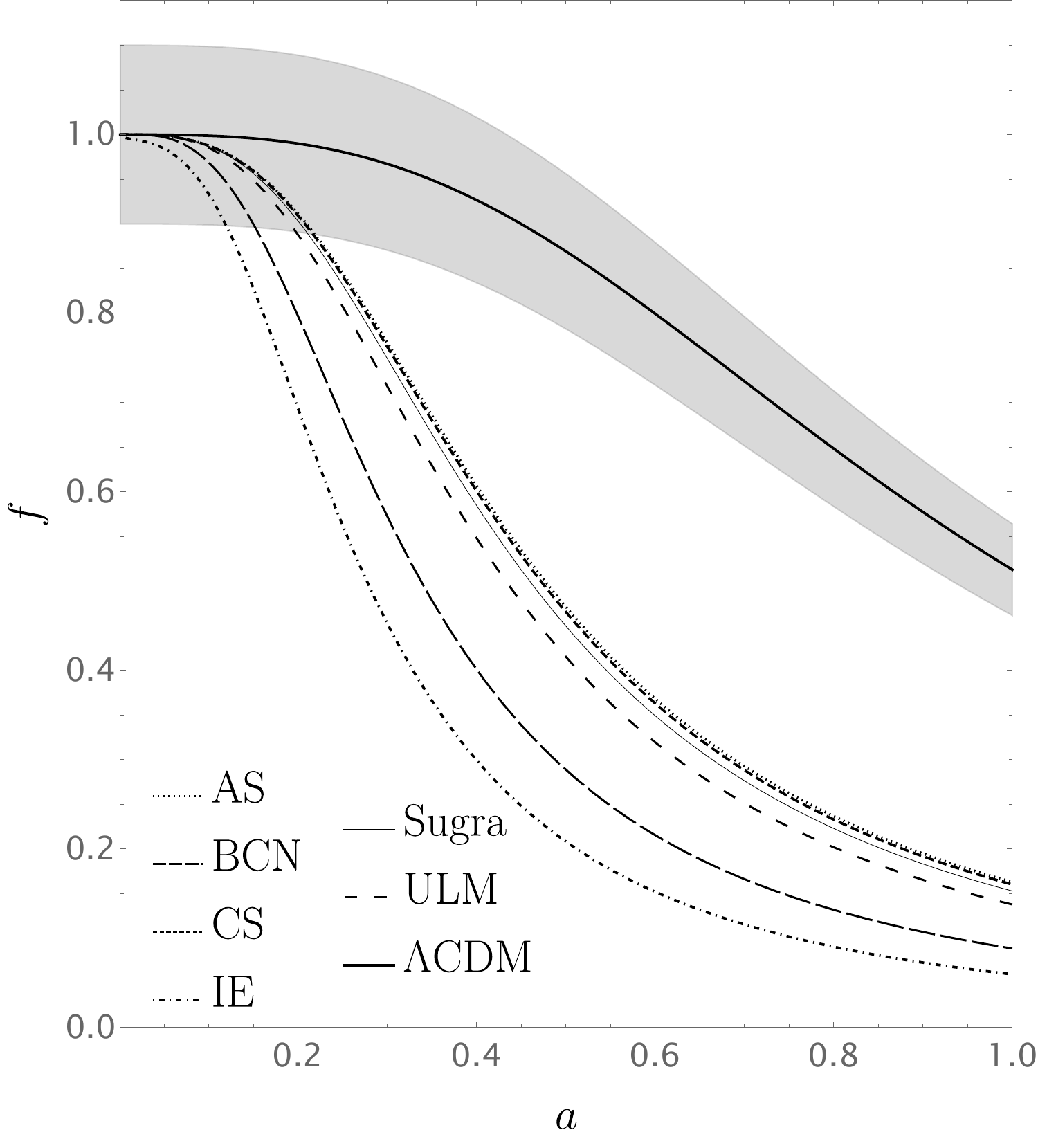}}
\hspace{2mm}
\subfigure[Growth index for the non-minimally coupled phantom fields.\label{PEDER2}]
{\includegraphics[height=0.52\hsize,clip]{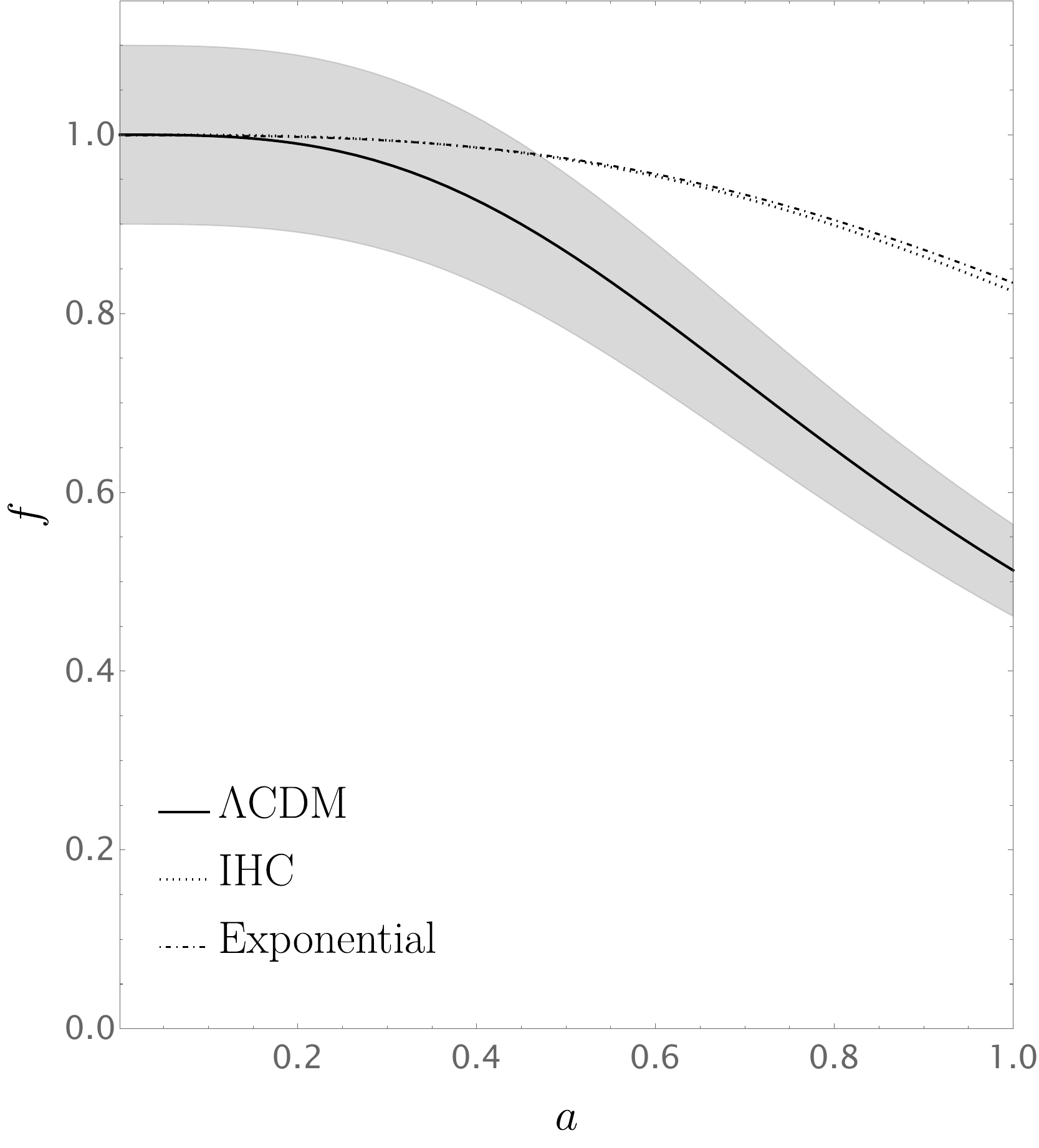}}
\caption{Growth index for the non-minimally coupled standard scalar field compared to the $\Lambda$CDM model with $\xi=-10^{-1}$ for IE, $\xi=10^{-2}$ for other quintessence models and $\xi=-0.5$ for phantom fields. The boundary conditions are imposed as $v^{2}_{\rm in}=0.999$, $u_{\rm in}\in\left[10^{-2},10^{-1}\right]$, $y_{\rm in}\in\left[10^{-2},10^{-1}\right]$, $N_{i}=\ln a_{\rm LSS}$ and $N_{f}=0$. The behavior does not change if we consider the alternative scalar field. The grey band represents $\pm 10 \%$ departures from $\Lambda$CDM.}\label{PEDERALL}
\end{figure*}

\begin{figure*}
\centering
\subfigure[Growth index for teleparallel and symmetric-teleparellel quintessence.\label{PEDET}]
{\includegraphics[height=0.52\hsize,clip]{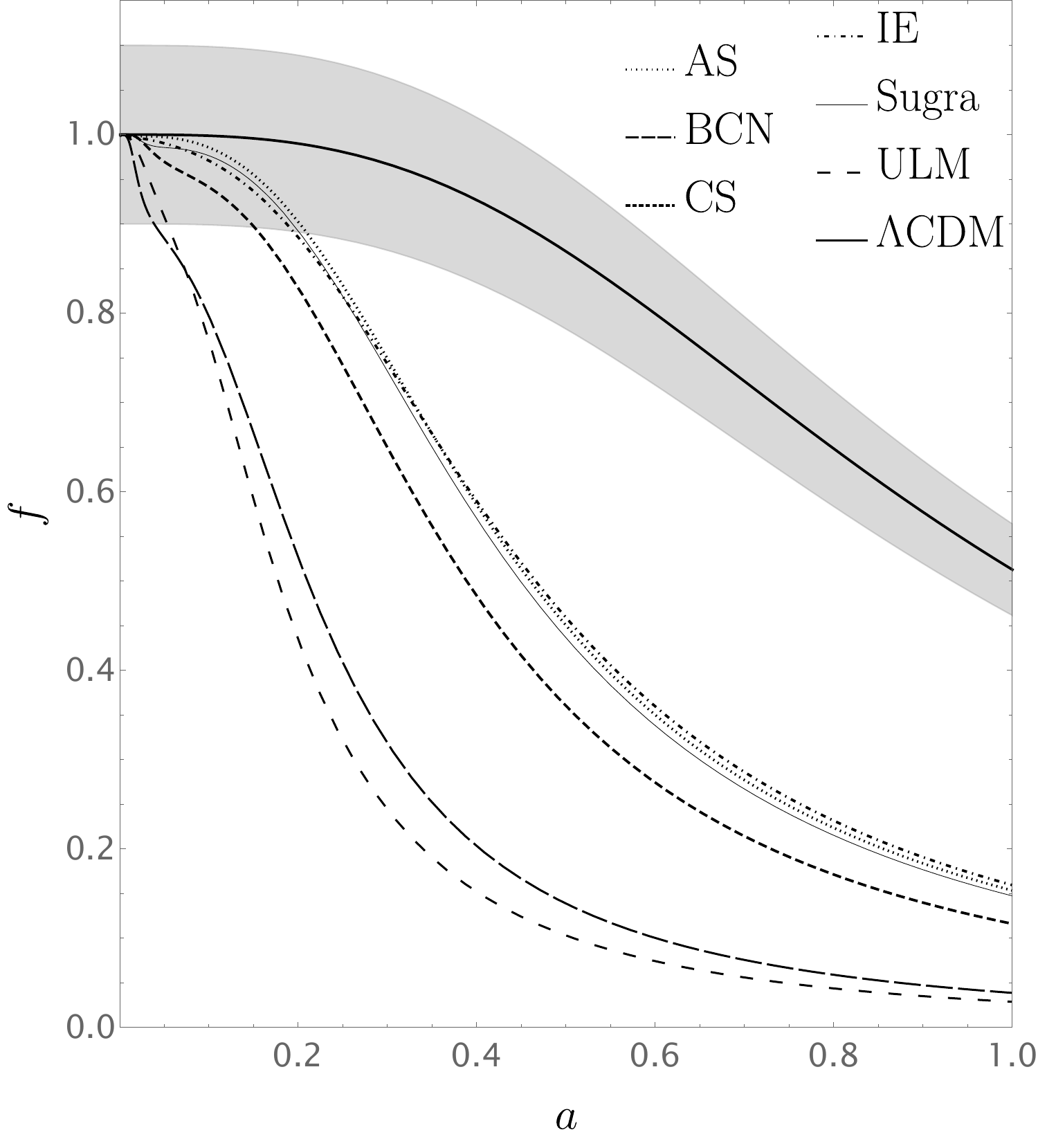}}
\hspace{2mm}
\subfigure[Growth index for teleparallel and symmetric-teleparellel phantom fields.\label{PEDET2}]
{\includegraphics[height=0.52\hsize,clip]{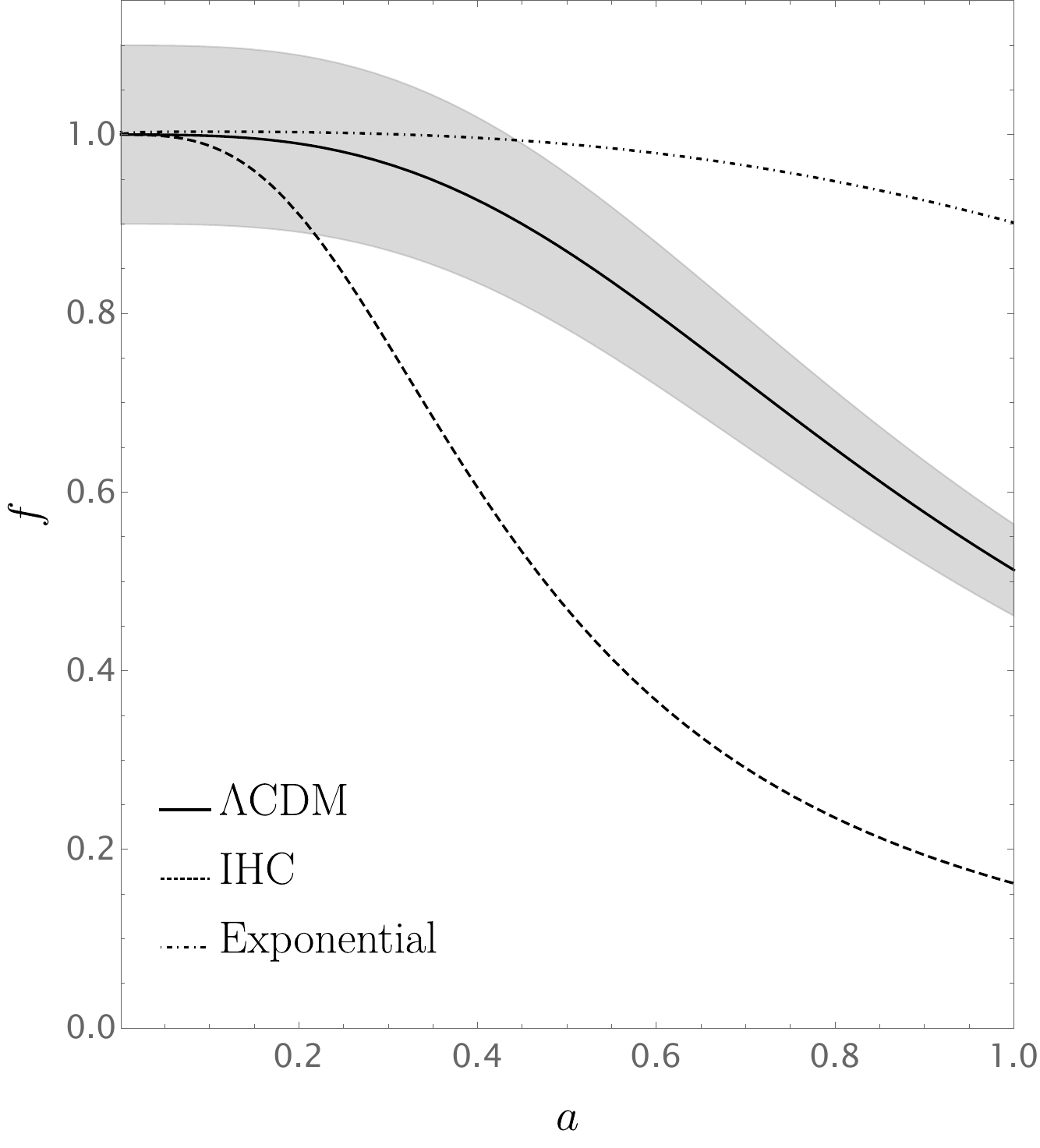}}
\caption{Growth index for teleparallel and symmetric-teleparallel standard scalar field compared to the $\Lambda$CDM model with $\xi \in \left[-10^{-3},-10^{-1}\right]$ for quintessence and $\xi \in \left[10^{-2},10^{-1}\right]$ for phantom fields. The boundary conditions are imposed as $v^{2}_{\rm in}=0.999$, $u_{\rm in}\in\left[10^{-2},1\right]$, $y_{\rm in}\in\left[10^{-2},10^{-1}\right]$, $N_{i}=\ln a_{\rm LSS}$ and $N_{f}=0$. The behavior does not change if we consider the alternative
scalar field. The grey band represents $\pm 10 \%$ departures from $\Lambda$CDM.}\label{PEDETALL}
\end{figure*}

\begin{figure*}
\centering
\subfigure[Growth index for the minimally coupled quasiquintessence.\label{PEDE}]
{\includegraphics[height=0.52\hsize,clip]{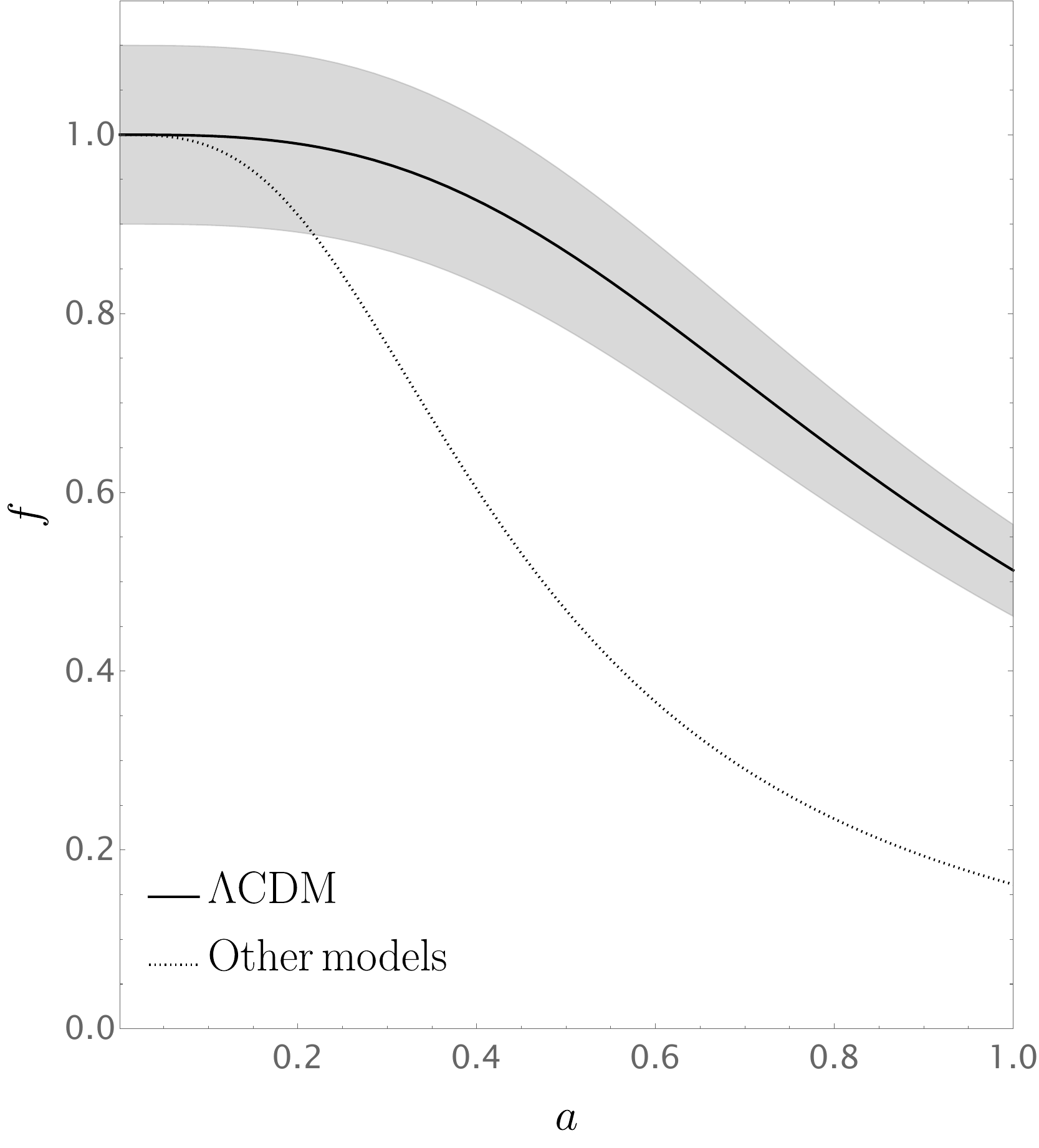}}
\hspace{2mm}
\subfigure[Growth index for the minimally coupled quasiphantom fields.\label{PEDE2}]
{\includegraphics[height=0.522\hsize,clip]{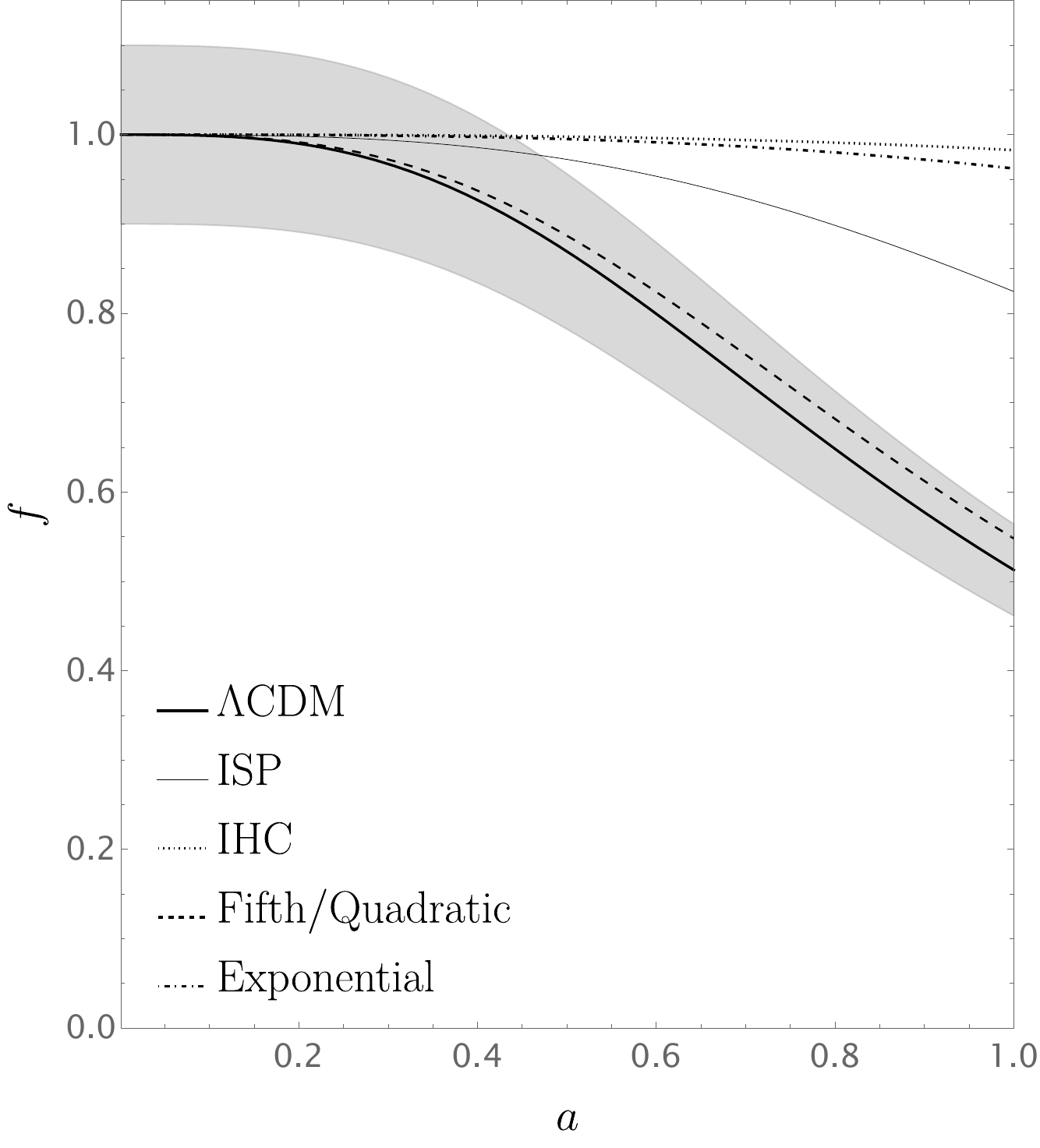}}
\caption{Growth index for the minimally coupled alternative scalar field compared to the $\Lambda$CDM model. The boundary conditions are imposed as $v^{2}_{\rm in}=0.999$, $u_{\rm in}\in\left[10^{-1},1\right]$, $y_{\rm in}\in\left[10^{-2},10^{-1}\right]$, $N_{i}=\ln a_{\rm LSS}$ and $N_{f}=0$. The grey band represents $\pm 10 \%$ departures from $\Lambda$CDM. }\label{PEDEALL}
\end{figure*}


\section{Outlooks and perspectives}\label{Sec6}

Motivated by the revived interest in evolving dark energy models, as recently claimed by the DESI collaboration, we reconsidered whether minimal and non-minimal coupled scalar field dark energy models can suitably behave from a dynamical viewpoint. Particularly, the DESI collaboration has pointed out that an evolving dark energy contribution can be due to a unknown scalar field dynamics that, only in the simplest scenario, reduces to an effective $w_0w_a$CDM model \cite{DESI:2024mwx}.

To this end, we computed the stability and the dynamical systems associated with the most popular forms of dark energy models, characterized by viable scalar field potentials.

Moreover, to extend the findings indicated by DESI, we did not limit our analysis to quintessence only, but considered phantom regimes, as well as alternative field representations, dubbed quasiquintessence and quasiphantom pictures, conceptually derived as possible generalizations of K-essence models, but with the great advantage to furnish an identically zero sound speed, namely behaving as dust-like fluids.

We provided, for the alternative scalar field representations, robust arguments in their favor, remarking their use to guarantee inflationary stability, or in erasing the excess of cosmological constant contribution throughout a metastable phase.

In addition, to check the goodness of each dark energy model, we ensured their suitability switching the theoretical background to alternatives to general relativity. Hence, we focused on general relativity first and then on its equivalent versions, offered by the teleparallel and symmetric-teleparallel formalism, by using Lagrangian densities, $R, T$ and $Q$, respectively.

More in detail, for each gravitational background and for the set of dark energy potentials, we investigated minimal and non-minimal couplings between the scalar field and the gravity sector, adopting Yukawa-like interacting terms.

In so doing, we checked the main consequences of coupling dark energy with gravity, under the form of curvature, torsion or non-metricity and, moreover, in the case of non-minimal coupling, while the limiting case, $\xi\rightarrow 0$, leading to minimal coupling is also studied.

Nevertheless, even in the minimally coupled scenario, we developed a general approach to study the stability of the alternative scalar field. In particular, considering quasiquintessence, we discovered that there exists a critical point where dark energy exhibits dust-like characteristics. This is a non-hyperbolic point since one of the eigenvalues is zero, so we do not infer the stability limiting to the linear analysis. In this respect, we applied the center manifold theory and numerical methods, and, fixing free parameters, we conclude that it is an attractor for AS, BCN and ULM potentials, suggesting that these potentials in the context of quasiquintessence determine a unified dark energy-dark matter fluid.

In addition, another critical point can be an attractor under specific conditions. This point is also given in the approach that generalizes stability for the standard scalar field, and it describes a scenario in which the universe is fully dominated by dark energy in the form of a bare cosmological constant, i.e., $w_{\phi}=-1$.

We extended the generalize method to study the stability in the context of non-minimal coupling scenarios, without expliciting the relation between $u$ and $\gamma$. Here, we analyzed the standard and alternative forms of the scalar field, both yielding the same critical points, contrary to the minimal coupled framework. This result indicates that the presence of coupling tends to hide the differences between these two scalar field descriptions, leading only a subtle differences in the form of eigenvalues. This finding confirms the results obtained for non-minimal couplings in inflationary regimes, suggesting that possible non-minimal couplings may be under different forms than Yukawa-like, but are in tension with the Higgs inflation and the Starobinky potential schemes, where, instead, the Yukawa-like term appears essential.

Moreover, with the given set of variables, all the critical points are non-hyperbolic, requiring a numerical approach to determine whether a critical point with three eigenvalues having negative real parts is an attractor. In any case, all points candidate as attractors lead to a universe dominated by dark energy under the form of a cosmological constant, and the unified dark energy-dark matter framework found in the minimimally coupled scenario is lost.

In the non-minimal usual general relativity framework, the phantom field power-law potentials are excluded from the analysis for both types of scalar fields, as they are ruled out by our computation.

Further, we analyzed teleparallel and symmetric-teleparallel dark energy within coincident gauge at the same time, by virtue of the mathematical degeneracy between the two frameworks.

Here, we derived again that the standard and alternative scalar field descriptions give the same results, leading to attractor points where dark energy dominates the universe with $w_{\phi}=-1$. Notably, in the case of teleparallel and symmetric-teleparallel dark energy, only for the ISP potential we did not find the correct conditions to determine the behavior of the critical point.

Thus, we conclude that the quintessence and quasiquintessence models emerge as the most promising dark energy candidates, as they exhibit attractor points across all the gravity scenarios considered. Additionally, our analysis of the growth of matter perturbations at early times reveals that, among all the gravity frameworks studied, only the quasiphantom field minimally coupled to gravity with fifth power and quadratic potentials remains within $\pm 10\%$ deviations from $\Lambda$CDM.

In conclusion, we observe that:
\begin{itemize}
    \item[-] Quasiquintessence provides two interesting critical points in minimally coupled cosmology. One showing a unified dark energy-dark matter scenario, whereas the other leading to the standard framework, where dark energy dominates the universe under the form of a cosmological constant.
    \item[-] The stability analysis in non-minimally coupled dark energy framework confirms that phantom fields are disfavored compared to quintessence, and this is confirmed for quasiphantom field too.
    \item[-] For teleparallel and symmetric-teleparallel dark energy models, phantom fields are not fully-excluded, suggesting either the need for further investigation into the validity of Yukawa-like coupling in these gravity scenarios or possible differences among the background gravity theories.
    \item[-] Regarding matter perturbations, all the models analyzed in non-minimally coupled scenarios show disagreement with the $\Lambda$CDM model. In contrast, in the minimally coupled framework, only the quasiphantom field represented by the fifth power and quadratic potentials aligns with the concordance paradigm.
\end{itemize}

Looking ahead, we will single out the best potentials exploring their consequences in various non-minimal coupling settings, utilizing the new releases from the DESI collaboration and investigating the growth of matter perturbations. Moreover, it would be interesting to study the same in the context where spatial curvature is not set to zero, to check if its influence is relevant. Last but not least, our approach would be useful to clarify whether these potentials can be even generalized for early dark energy contexts, in view of a possible resolution of the cosmological tensions \cite{Poulin:2018cxd, Giare:2024akf}.

\section*{Acknowledgements}
YC acknowledges the Brera National Institute of Astrophysics (INAF) for financial support and Andronikos Paliathanasis for discussions on teleparallel and symmetric-teleparallel theories. OL is in debit with Eoin Ó Colgáin and M. M. Sheikh-Jabbari for fruitful discussions on the topic of dark energy evolution and stability. He also expresses his gratitude to Marco Muccino for interesting debates on dark energy scalar field representation. The authors sincerely acknowledge Rocco D'Agostino for the his support in choosing the best priors adopted for the numerical part of this manuscript.

\bibliographystyle{ieeetr}

\newpage
\appendix

\section{Center manifold theory}\label{CMT}

The non-hyperbolicity of a critical point arises when the Jacobian matrix, evaluated at that point, yields null eigenvalues. In such cases, the stability of the point cannot be determined using linear theory if all the non-zero eigenvalues have negative real parts. Thus, alternative techniques are needful, such as center manifold theory. This approach reduces the dimensionality of the system near the critical point, allowing the stability of the reduced system to be analyzed \cite{Boehmer:2011tp}. Below, we outline how the center manifold theory is applied throughout the text, following Ref. \cite{Das:2019ixt}.

\begin{itemize}
    \item[-] We first rewrite the system in terms of new variables by shifting the critical point to the origin of phase-space.

\item[-] Second, the dynamical system is split into a linear and a nonlinear part, say
\begin{eqnarray}
        & q' = Aq + f(q, p), \label{eq:CMT1} \\
        & p' = Bp + g(q, p), \label{eq:CMT2}
\end{eqnarray}

where $(q, p) \in \mathbb{R}^c \times \mathbb{R}^s$, whereas the functions $f$ and $g$ satisfy the conditions
\begin{eqnarray}
     & f(0, 0) = 0, \quad Df(0, 0) = 0, \\
     & g(0, 0) = 0, \quad Dg(0, 0) = 0.
 \end{eqnarray}
 Here, the critical point is at the origin, with $Df$ denoting the matrix of first derivatives of $f$. The matrix $A$ is a $c \times c$ matrix whose eigenvalues have zero real parts, while $B$ is an $s \times s$ matrix with eigenvalues having negative real parts.

 If the system is not in this appropriate form, a further change of variables is performed, determining the eigenvectors and eigenvalues of the Jacobian matrix.

 \item[-] Next, we introduce a function $h(q)$ and expand it in a Taylor series around the origin as $h(q) = aq^2 + bq^3 + \mathcal{O}(q^4)$. The coefficients $a$ and $b$ are determined by solving the quasilinear partial differential equation
 \begin{widetext}
    \begin{equation}
        Dh(q)(Aq + f(q, h(q))) - Bh(q) - g(q, h(q)) = 0, \label{eq:quasilinear}
    \end{equation}
 \end{widetext}
    where $h(0)=Dh(0)=0$.

\item[-] Once the coefficients are determined, we can analyze the dynamics of the original system, restricted to the center manifold, by writing
    \begin{equation}
        q'=Aq+f(q,h(q)).
    \end{equation}
 If at least one of the coefficients is non-zero, this equation reduces to $q'=\upsilon q^n$, where $\upsilon$ is a constant, and $n$ is a positive integer representing the lowest order in the expansion. If $\upsilon<0$ and $n$ is odd, the system is stable, and the critical point is an attractor. Otherwise, the critical point is unstable.
\end{itemize}
Now, we apply this technique to the critical point $P_4=(0, 1, 0)$, identified for the alternative scalar field model minimally coupled to gravity. We first shift the coordinates as $x=X$, $y-1=Y$, and $\lambda=\Lambda$, and rewrite the system in terms of these new variables. Thus, the dynamical system becomes
\begin{equation}
\begin{cases}
    X'=-\frac{3}{2}X (Y - 1)^2 + \epsilon\sqrt{\frac{3}{2}}\Lambda (Y-1)^2,& \\
    Y'=-\sqrt{\frac{3}{2}} \lambda X (Y - 1) + \frac{3}{2}(Y-1)\left(1-(Y-1)^2\right),& \\
    \Lambda'=-\sqrt{6}\Lambda^2\left(\Gamma(\Lambda)-1\right)X.&
\end{cases}
\end{equation}
We compute the eigenvector matrix of the new system's Jacobian and introduce a new set of coordinates as follows
\begin{equation}
    \begin{pmatrix}
        v \\
        w \\
        z
    \end{pmatrix}
    = \begin{pmatrix}
        0 & 1 & 0 \\
        1 & 0 & -\sqrt{\frac{2}{3}} \\
        0 & 0 & 1
    \end{pmatrix}
    \begin{pmatrix}
        X \\
        Y \\
        \Lambda
    \end{pmatrix}.
\end{equation}
Thus, with $v=Y$, $w=X-\sqrt{\frac{2}{3}}\Lambda$, and $z=\Lambda$, we rewrite the system in linear and nonlinear parts
\begin{equation}
    \begin{pmatrix}
        v' \\
        w' \\
        z'
    \end{pmatrix}
    = \begin{pmatrix}
        -3 & 0 & 0 \\
        0 & -\frac{3}{2} & 0 \\
        0 & 0 & 0
    \end{pmatrix}
    \begin{pmatrix}
        v \\
        w \\
        z
    \end{pmatrix}
    + \begin{pmatrix}
        \text{nonlinear part}
    \end{pmatrix},
\end{equation}
where the matrix represents the eigenvalue matrix.

At this stage, we compare our system with the dynamical system defined by Eqs. \eqref{eq:CMT1} and \eqref{eq:CMT2}. We generalize it by setting $q = z$ and $p=(v, w)$, and apply center manifold theory. This gives
\begin{widetext}
\begin{eqnarray}
    &A = 0, \quad B = \begin{pmatrix} -3 & 0 \\ 0 & -\frac{3}{2} \end{pmatrix}, \\
&f = \sqrt{6}z^2\left(\Gamma(z) - 1\right)(w + \sqrt{\frac{3}{2}}z), \\
    &g = \begin{pmatrix}
        \frac{1}{2} \left(-3(v - 3) v^2 - \sqrt{6}(v - 1) wz - 3(v - 1)z^2\right) \\
        \frac{1}{4} \left(-6w \left((v - 2)v + 2z^{\frac{3}{2}}\right) - \sqrt{6}z \left(5(v - 1)^2 + 6z^{\frac{3}{2}}\right)\right)
    \end{pmatrix}.
\end{eqnarray}
\end{widetext}

To analyze the dynamics of the system, we replace $p$ with $h$, assumed under the form,
\begin{equation}
    h \equiv \begin{pmatrix}
        a_2 z^2 + a_3 z^3 + \mathcal{O}(z^4) \\
        b_2 z^2 + b_3 z^3 + \mathcal{O}(z^4)
    \end{pmatrix}.
\end{equation}
Solving Eq. \eqref{eq:quasilinear}, we find

\begin{equation}
    a_3= b_2=0,\hspace{2mm}a_2=\frac{1}{2},\hspace{2mm}b_3=\sqrt{\frac{3}{2}}-\sqrt{6}\epsilon.
\end{equation}

The system restricted to the center manifold is then
\begin{equation}
    z'=-2 z^3(\Gamma(0)-1)+\mathcal{O}(z^4),
\end{equation}
indicating that the critical point $P_4$ is stable if $\Gamma(0)>1$, and unstable if $\Gamma(0)<1$.

\section{Phase-space numerical analysis for dark energy non-minimally coupled to gravity}

In this appendix, we present the phase-space analysis for non-minimally coupled scalar fields. Figs.~\ref{StRmodels} and~\ref{StTQmodels} are obtained by fixing the free parameters of the potentials and the coupling constant.

\begin{figure*}
\centering
\subfigure[Sugra with $\chi=\gamma=1$ and $\xi=\frac{1}{10}$.\label{StRSugra}]
{\includegraphics[height=0.4\hsize,clip]{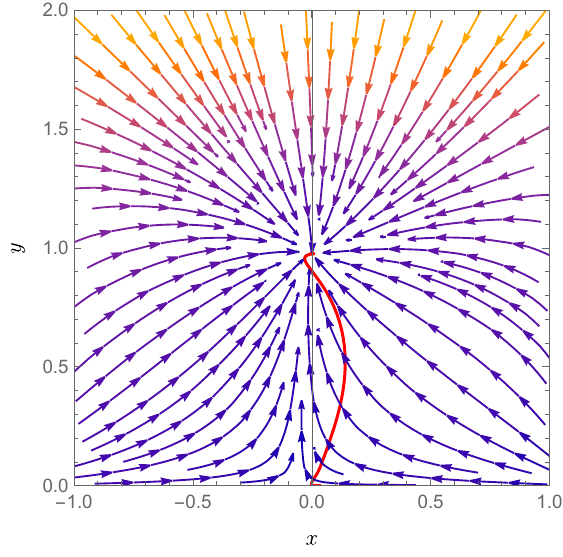}}
\subfigure[BCN with $l=6$, $m=\frac{1}{2}$ and $\xi=\frac{1}{2}$.\label{StRBCN}]
{\includegraphics[height=0.4\hsize,clip]{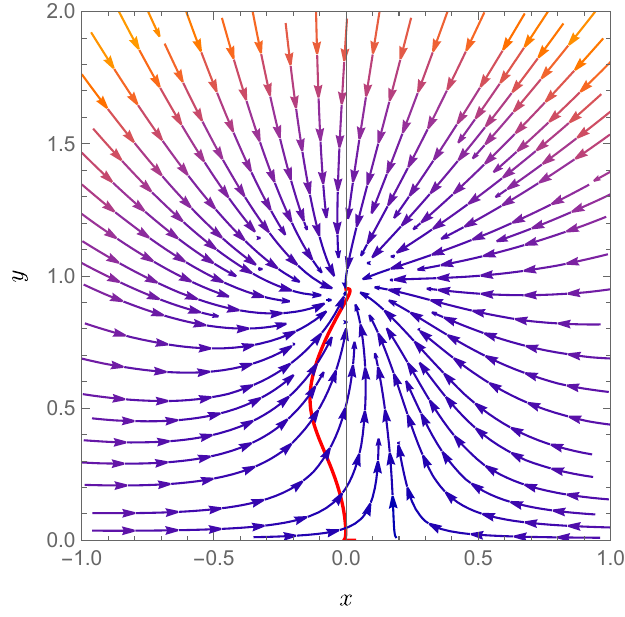}}
\subfigure[AS with $A=B=\mu=\frac{1}{6}$ and $\xi=\frac{1}{10}$.\label{StRAS}]
{\includegraphics[height=0.4\hsize,clip]{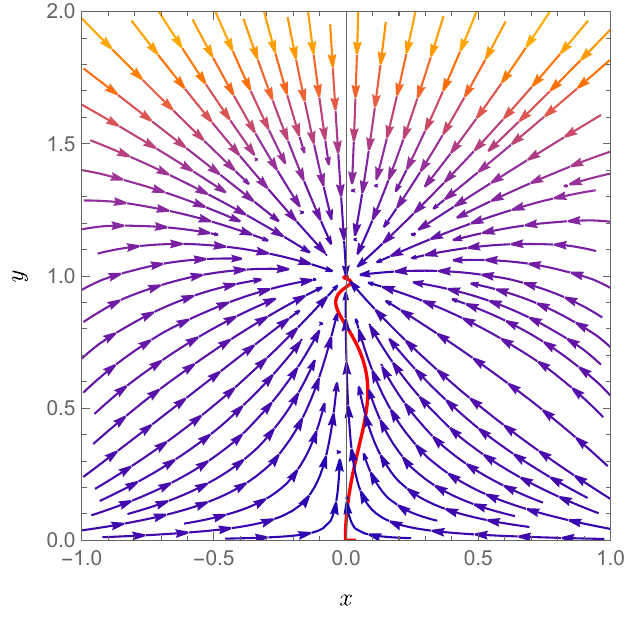}}
\subfigure[ULM with $n=-\frac{1}{2}$, $\zeta=\frac{1}{2}$ and
 $\xi=\frac{1}{2}$.\label{StRULM}]
{\includegraphics[height=0.4\hsize,clip]{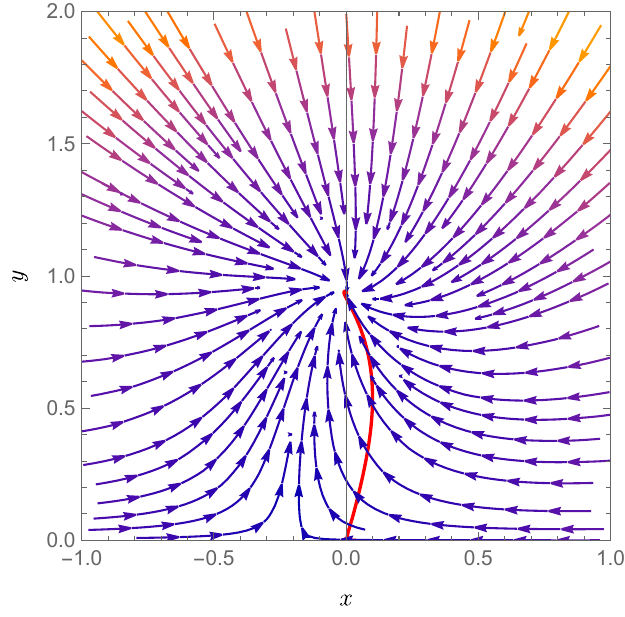}}
\end{figure*}
\begin{figure*}
\centering
\subfigure[IE with $\xi=\frac{1}{6}$.\label{StRExp-1}]
{\includegraphics[height=0.4\hsize,clip]{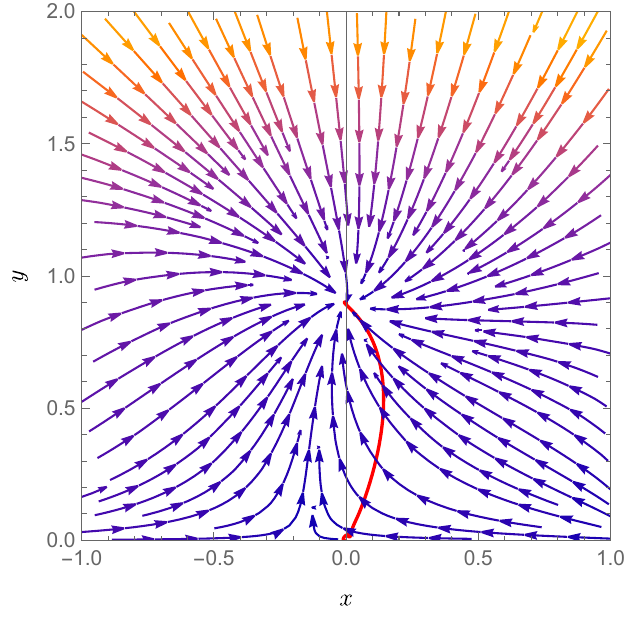}}
\subfigure[CS with $\tau=1$ and $\xi=1$.\label{StRCS}]
{\includegraphics[height=0.4\hsize,clip]{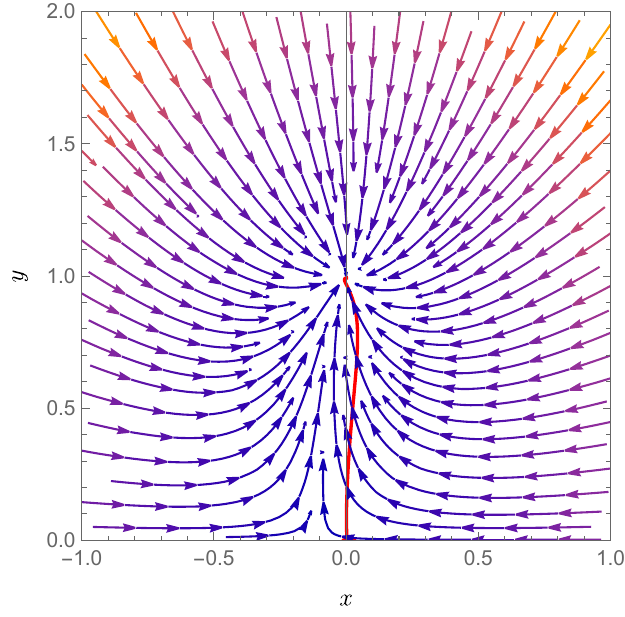}}
\subfigure[Exponential with $\beta=\frac{1}{10}$ and $\xi=\frac{1}{2}$.\label{StRPhantExp}]
{\includegraphics[height=0.4\hsize,clip]{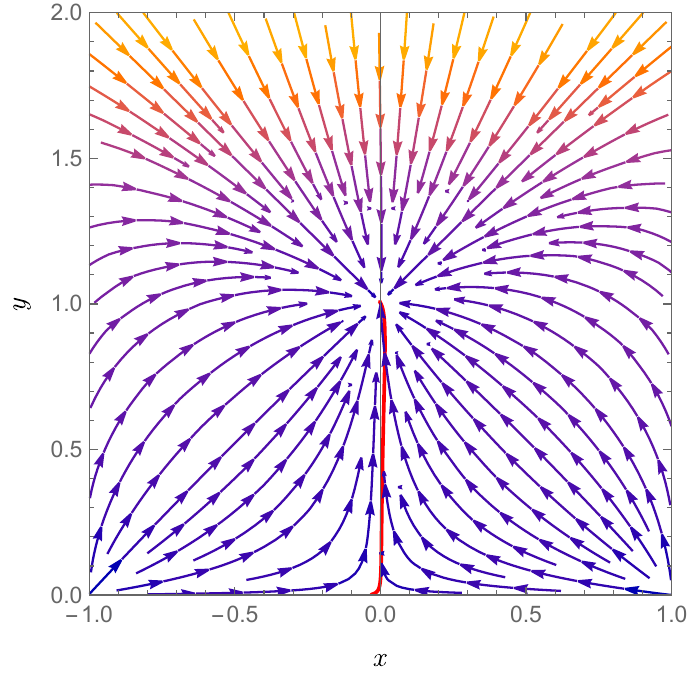}}
\subfigure[IHC with $\alpha=2$ and $\xi=\frac{1}{6}$.\label{StRIHC}]
{\includegraphics[height=0.4\hsize,clip]{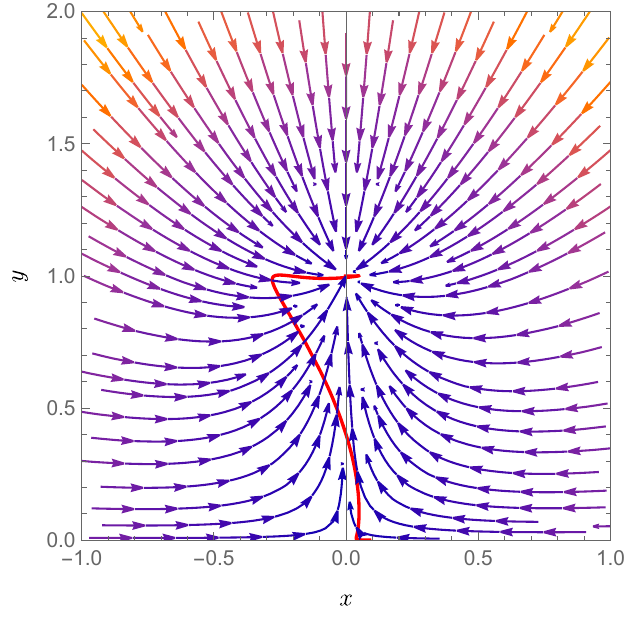}}
\caption{Phase-space trajectories on the $x-y$ plane for non-minimally coupled standard scalar field models. The coupling constant and initial conditions are also valid for the alternative scalar field description. Red lines indicate the solutions of the dynamical systems.}
\label{StRmodels}
\end{figure*}

\begin{figure*}
\centering
\subfigure[Sugra with $\chi=1$, $\gamma=\frac{1}{2}$ and $\xi=-\frac{1}{2}$.\label{StTQSugra}]
{\includegraphics[height=0.4\hsize,clip]{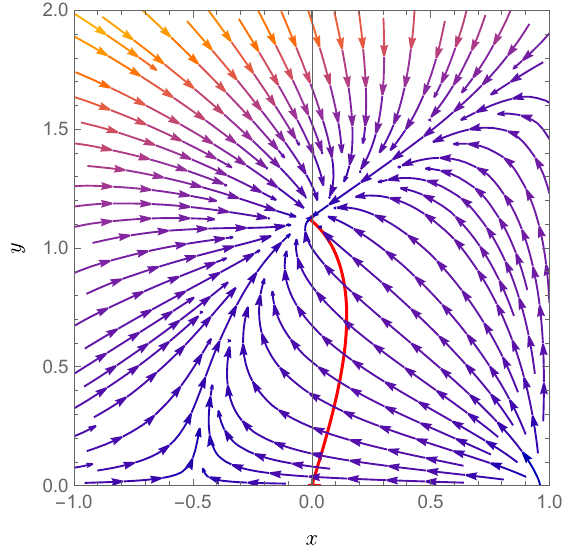}}
\subfigure[BCN with $l=6$, $m=\frac{1}{2}$ ad $\xi=-1$.\label{StTQBCN}]
{\includegraphics[height=0.4\hsize,clip]{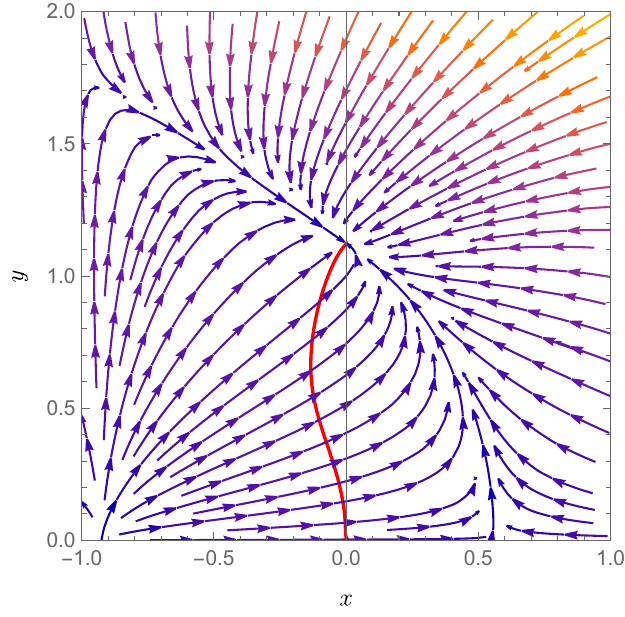}}
\subfigure[AS with $A=B=\mu=\frac{1}{6}$ and $\xi=-\frac{1}{2}$.\label{StTQAS}]
{\includegraphics[height=0.4\hsize,clip]{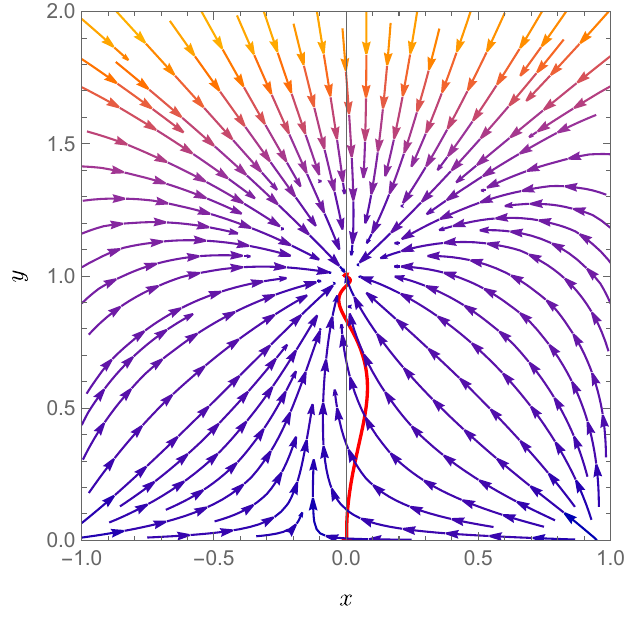}}
\subfigure[ULM with $n=-\frac{1}{2}$, $\zeta=\frac{1}{2}$ and $\xi=-\frac{1}{2}$.\label{StTQULM}]
{\includegraphics[height=0.4\hsize,clip]{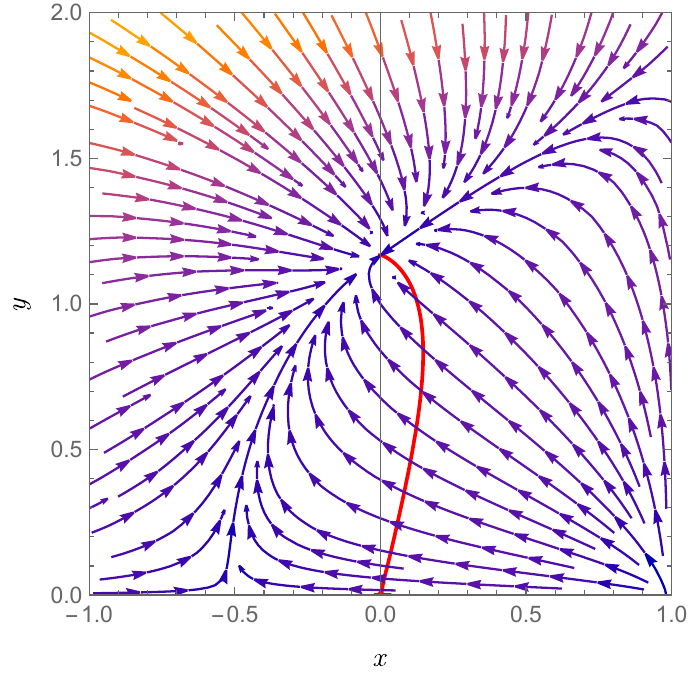}}
\subfigure[IE with $\xi=-\frac{1}{10}$.\label{StTQIE}]
{\includegraphics[height=0.4\hsize,clip]{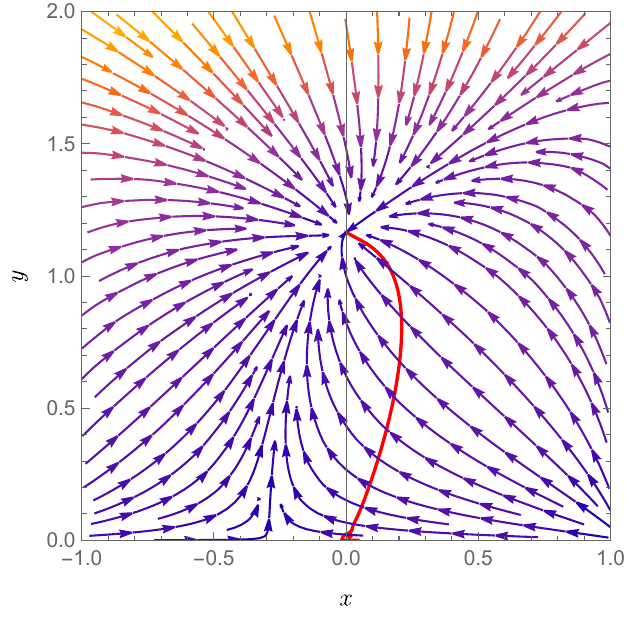}}
\subfigure[CS with $\tau=1$ and $\xi=-1$.\label{StTQCS}]
{\includegraphics[height=0.4\hsize,clip]{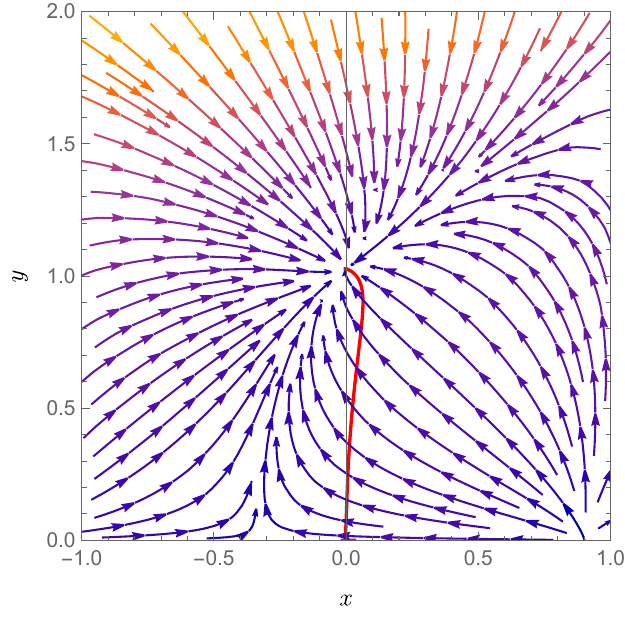}}
\end{figure*}
\begin{figure*}
\centering
\subfigure[Fifth power with $\xi=1$. \label{StTQ5th}]
{\includegraphics[height=0.4\hsize,clip]{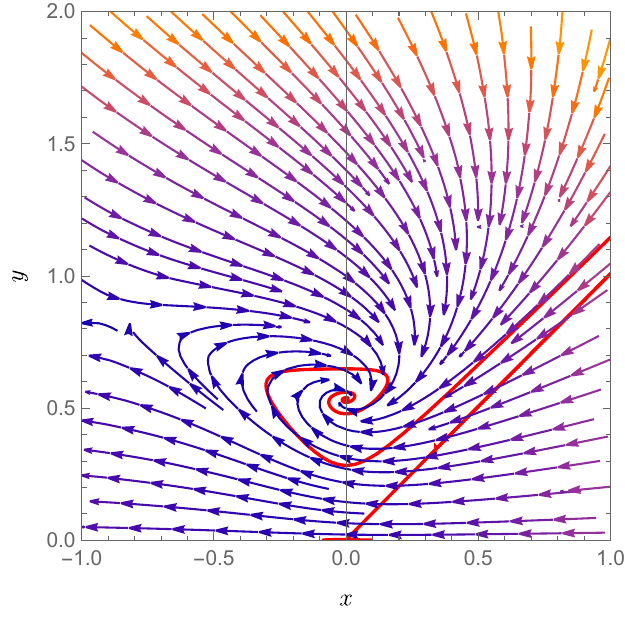}}
\subfigure[Quadratic with $\xi=1$.\label{StTQ2th}]
{\includegraphics[height=0.4\hsize,clip]{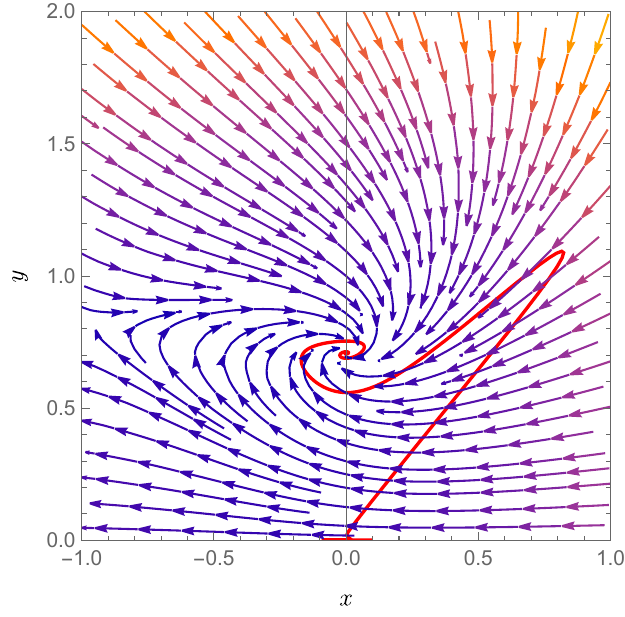}}
\subfigure[Exponential with $\beta=1$ and $\xi=\frac{1}{2}$.\label{StTQE}]
{\includegraphics[height=0.4\hsize,clip]{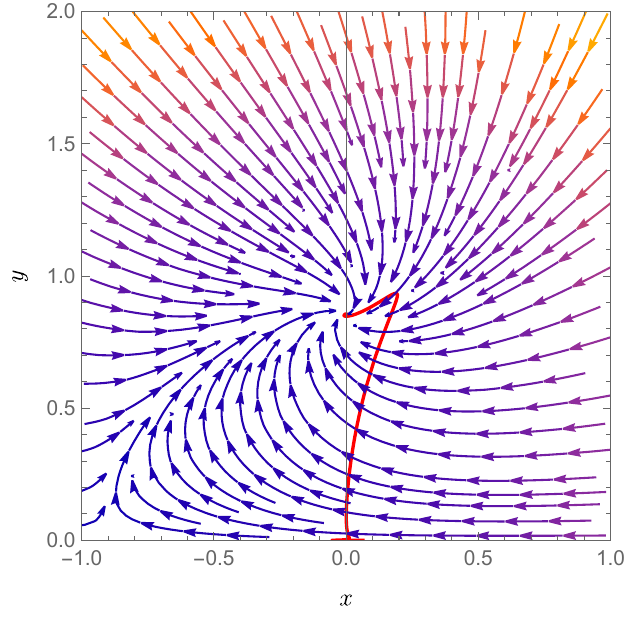}}
\subfigure[IHC with $\alpha=1$ and $\xi=1$.\label{StTQIHC}]
{\includegraphics[height=0.4\hsize,clip]{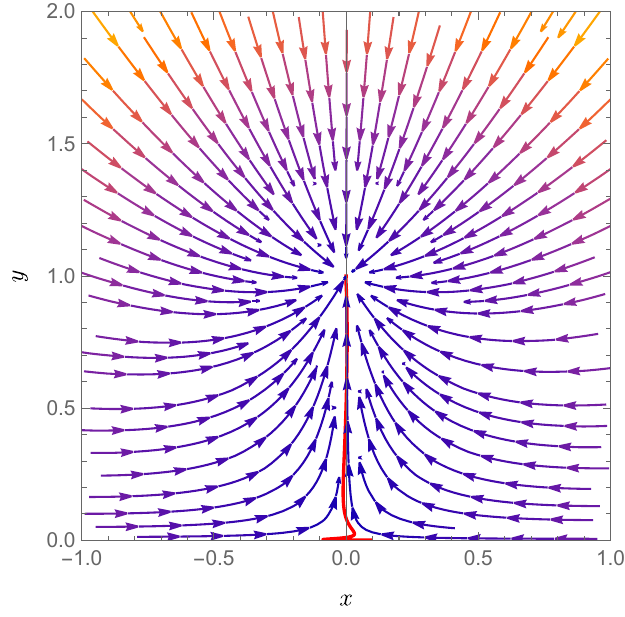}}
\caption{Phase-space trajectories on the $x-y$ plane for teleparallel and symmetric-teleparallel standard scalar field. The coupling constant and initial conditions are chosen to be valid for both the standard and alternative scalar field descriptions. Red lines indicate the solutions of the dynamical systems.}
\label{StTQmodels}
\end{figure*}

All the critical points depicted in Figs.~\ref{StRmodels} and~\ref{StTQmodels} are attractor points, where the dynamics of the systems reach a stable state.

In the context of non-minimally coupled dark energy in Einstein's theory, phantom fields are mostly excluded from the analysis, allowing us to identify attractor points only for the exponential and IHC potentials.

In contrast, for teleparallel and symmetric-teleparallel dark energy, phantom field scenario is almost reaffirmed, say it cannot be ruled out,  with the only exception of the inverse power law potential.

Thus, using these variables, quintessence and quasiquintessence turn out to be favored at least from a numerical viewpoint, however still degenerating between them.

Specifically, setting an appropriate coupling constant, standard and alternative scalar field descriptions give the same outcomes.

Finally, it is remarkable to stress that the coupling hides the dust-like characteristics of the alternative scalar field, dominating over it through the squared term coupled to $U$.

\end{document}